\documentclass[lettersize,journal]{IEEEtran}
\usepackage{amsmath,amsfonts}
\usepackage{algorithmic}
\usepackage{algorithm}
\usepackage{array}
\usepackage[caption=false,font=normalsize,labelfont=sf,textfont=sf]{subfig}
\usepackage{textcomp}
\usepackage{stfloats}
\usepackage{url}
\usepackage{verbatim}
\usepackage{graphicx}
\usepackage{cite}
\usepackage{enumitem}

\newcommand{\hide}[1]{}
\newcommand{\vpara}[1]{\vspace{0.2cm}\noindent\textbf{#1 }}

\usepackage[misc]{ifsym}

\usepackage{booktabs}
\usepackage[table,xcdraw]{xcolor}
\usepackage{textcomp} 

\begin{document}

\title{LGB: Language Model and Graph Neural Network-Driven Social Bot Detection}

\author{Ming Zhou,
Dan Zhang,
Yuandong Wang,
Yangli-ao Geng,
Yuxiao Dong,
and Jie Tang,~\IEEEmembership{Fellow, IEEE}
\IEEEcompsocitemizethanks{\IEEEcompsocthanksitem Ming Zhou, Dan Zhang, Yuandong Wang, Yuxiao Dong, and Jie Tang are with the Department of Computer Science and Technology, Tsinghua Univerisity, China. E-mail: zhou-m19@mails.tsinghua.edu.cn, jietang@tsinghua.edu.cn
\IEEEcompsocthanksitem Yangli-ao Geng is with the Department of Computer Science and Technology, Beijing Jiaotong University, China.
\IEEEcompsocthanksitem Corresponding author: Jie Tang.}
}

\markboth{Journal of \LaTeX\ Class Files,~Vol.~14, No.~8, August~2021}%
{Shell \MakeLowercase{\textit{et al.}}: A Sample Article Using IEEEtran.cls for IEEE Journals}

\maketitle

\begin{abstract}
Malicious social bots achieve their malicious purposes by spreading misinformation and inciting social public opinion, seriously endangering social security, making their detection a critical concern. Recently, graph-based bot detection methods have achieved state-of-the-art (SOTA) performance. However, our research finds many isolated and poorly linked nodes in social networks, as shown in Fig.~\ref{fig:Figure1_data_show}, which graph-based methods cannot effectively detect. To address this problem, our research focuses on effectively utilizing node semantics and network structure to jointly detect sparsely linked nodes. Given the excellent performance of language models (LMs) in natural language understanding (NLU), we propose a novel social bot detection framework LGB, which consists of two main components: language model (LM) and graph neural network (GNN). Specifically, the social account information is first extracted into unified user textual sequences, which is then used to perform supervised fine-tuning (SFT) of the language model to improve its ability to understand social account semantics. Next, the semantically enriched node representation is fed into the pre-trained GNN to further enhance the node representation by aggregating information from neighbors. Finally, LGB fuses the information from both modalities to improve the detection performance of sparsely linked nodes. Extensive experiments on two real-world datasets demonstrate that LGB consistently outperforms state-of-the-art baseline models by up to $10.95\%$. LGB is already online: \url{https://botdetection.aminer.cn/robotmain}.
\end{abstract}


\section{Introduction}
\IEEEPARstart{A}{s} multimedia-rich social networks become deeply integrated into our daily lives, and their influence grows inevitable. Concurrently, the rapidly developing artificial intelligence (AI) technology has achieved remarkable success in various fields, alongside new challenges, notably the rise of malicious social bots. Social bots are automated agents that are fully or partially controlled by computer programs~\cite{social_bot_web}. These bots are evolving to think, speak, and interact in an increasingly human-like manner for malicious purposes. Over the last decade, such bots have been implicated in spreading misinformation and fake news, impacting public opinion and financial markets~\cite{shao2018spread},\cite{azzimonti2023social},\cite{cai2023network}. During the COVID-19 pandemic, social bots were found to contribute $9.27\%$ of tweets to discussions about the pandemic on Twitter/X, and studies show they successfully spread anger toward humans~\cite{shi2020social} and are involved in the generation and dissemination of false information about the COVID-19 vaccine~\cite{zhang2022social}. Recent research finds that social bots are widely used in network information warfare in the Russia-Ukraine war~\cite{smart2022istandwithputin,li2023influence}. For example, De Faveri et al.~\cite{de2023twitter} find that around $12\%$ of commentators on the Russia-Ukraine war during the $2022$ Italian general election were bots, and their analysis shows that bots influenced people's opinions by distorting the way war issues were treated. Furthermore, bots are involved in manipulating election outcomes to undermine regional security~\cite{woolley2016automating}. Specifically, they influence public opinion by distorting facts, spreading fake news, and attacking opponents, for example, the $2019$ Spanish general election~\cite{pastor2020spotting}, the $2016$ U.S. presidential election~\cite{luceri2019evolution}, the $2018$ U.S. midterm elections~\cite{luceri2019red}, etc. Elon Musk's proposed $\$44$ billion acquisition of Twitter was halted due to concerns over the prevalence of fake accounts~\cite{musk_fake_account_web}, highlighting the seriousness of the social bot problem. Social bots distort facts and manipulate public opinion by spreading false information, posing serious threats to financial security, health and epidemic prevention, social security, and world peace. Hence, there is a critical need for effective and reliable social bot detection techniques to ensure social safety and harmony.

\begin{figure}[t!]
\centering
\includegraphics[width=\linewidth]{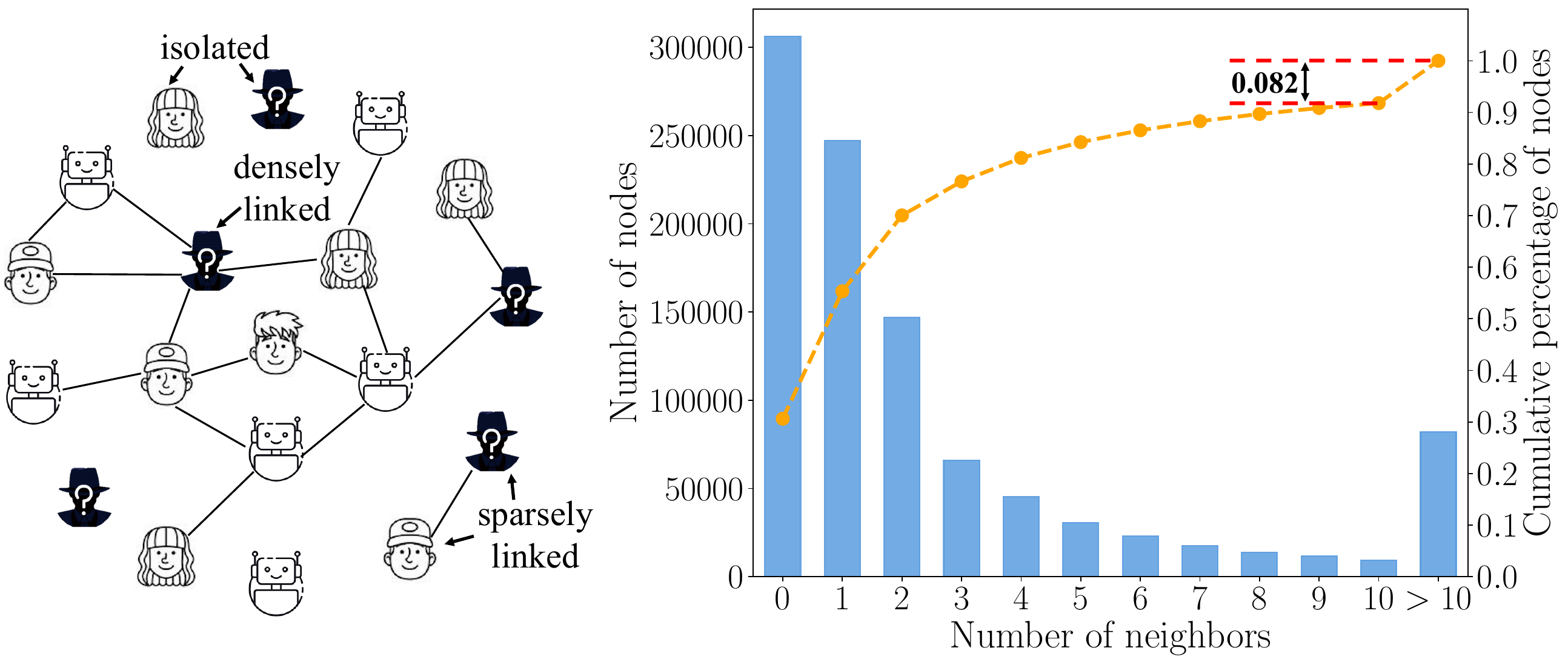}
\caption{{We conduct a Pareto analysis of the distribution of social relationships on TwiBot-22~\cite{feng2022twibot}, a real-world social network dataset (left), and find that there are a large number of isolated and poorly linked nodes in the social network (right). Specifically, isolated nodes account for as high as $30.62\%$ of all nodes, nodes with only one neighbor make up about $24.71\%$, but nodes with more than ten neighbors constitute only $8.2\%$.}} \label{fig:Figure1_data_show}
\vspace{-0.6cm}
\end{figure}

In the early days of the development of social bot detection technology, the primary approaches are \textbf{featured-based}. These methods construct user features from information such as user attributes~\cite{yang2020scalable}, user behaviors~\cite{sayyadiharikandeh2020detection,cai2017behavior}, and tweets~\cite{heidari2021empirical,wang2018social} based on statistical tools and expert knowledge. However, such methods have poor scalability and are easily attacked by feature forgery~\cite{cresci2020decade}, where bot developers modify features to evade detectors. To combat the dissemination of misinformation by social bots, \textbf{content-based} detection methods are proposed, where natural language processing (NLP) technologies are widely used to detect accounts by evaluating the authenticity and purpose of tweet content. For example, Wei Feng et al. employ bidirectional Long Short-term Memory (BiLSTM) to extract content features to detect bots~\cite{wei2019twitter}. Cai et al. adopt convolutional neural networks (CNNs) to obtain the features of tweets for bot detection~\cite{cai2017behavior}. However, the emergence of Large-scale Language Models (LLMs) in recent years is empowering social bots with stronger content creation capabilities. For example, OpenAI's newly released content classifier can only correctly identify $26\%$ of AI-written content~\cite{aiclassifierweb}. This new challenge is weakening the performance of content-based detection methods. 
Given that social bots mainly achieve their malicious purposes by spreading false information, inspired by the research finding that the strength~\cite{bakshy2012role} and structural diversity~\cite{ugander2012structural, zhang2013social} of social relationships play an important role in the spread of information, \textbf{graph-based} methods that detect accounts by modeling social relationships are proposed and have great promise in detecting bot group attacks~\cite{latah2020detection, zhou2023semi}.
For example, Zhou et al.~\cite{zhou2023detecting} propose a contrastive learning-based social bot detection approach CBD. However, our research reveals that social networks contain a significant number of isolated and sparsely linked nodes. Specifically, up to $30.62\%$ of nodes are isolated, and approximately $24.71\%$ have only one neighbor, as shown in Fig.~\ref{fig:Figure1_data_show}. For such nodes, the detection performance of traditional graph-based methods will decline, which greatly weakens the detector's ability to identify bots in the early and hidden stages. These bots will be quickly activated to establish links with humans when performing malicious tasks to spread false information and engage in malicious activities. Such bots are extremely harmful and difficult to detect using single-modality detection methods, posing significant challenges to the social bot detection task.

To effectively detect isolated and sparsely linked nodes, we propose LGB, a novel multimodal social bot detection framework, which combines the semantic understanding capabilities of language models (LMs) with the network structure extraction capabilities of graph neural networks (GNNs) to achieve cross-modal joint detection of social accounts. Specifically, first, social information such as user attributes, personal descriptions, and tweets of social accounts are extracted to form user text. Second, based on the user text, supervised fine-tuning is performed on the LM to improve its ability to understand social account information. Then, the semantically enhanced node representation is fed into GNN to further integrate network structure information. Finally, our model improves the detection performance of isolated and sparsely linked nodes by fusing the two modalities of text semantics and network structure. For LGB's system architecture, our framework adopts the design paradigm of online and offline dual systems to achieve better scalability. In the online system, we innovatively propose a smart feedback strategy to correct erroneous prediction results in time. These corrected results are fed back into the offline system for adjustment in the next round of model training.

\vspace{-0.2cm}
\vpara{Contributions:} In summary, the main contributions of this work include:
\vspace{-0.2cm}

\begin{itemize}[leftmargin=*,itemsep=0pt,parsep=0.5em,topsep=0.3em,partopsep=0.3em]
    \setlength{\itemsep}{0pt}
    \setlength{\parsep}{0pt}
    \setlength{\parskip}{0pt}
    \item By analyzing social network data, we find that approximately $55.34\%$ of nodes in the network are isolated or have only one neighbor, as shown in Fig.~\ref{fig:Figure1_data_show}. Traditional graph-based detection methods have difficulty in identifying these nodes. Considering the rich semantic information of social accounts and the social semantic knowledge learned by the LM during pre-training, we investigate the effectiveness of LM and GNN in the social bot detection task. We find that for isolated and sparsely linked nodes, the supervised fine-tuned LM can effectively detect them, whereas for densely linked nodes, GNN achieves better detection performance. Additionally, our structural analysis of social relationships reveals an intrinsic link between social relationship structure and bot probability, which proves the importance of structural information for account detection tasks. All these findings inspire us to fuse node semantics with network structure to improve detection performance.
    \item We propose LGB, a novel bot detection framework that combines the semantic understanding ability of LMs and the network structure extraction ability of GNNs to achieve cross-modal joint social account detection. Moreover, at the system architecture design level, the design paradigm based on online and offline dual systems is adopted to improve the system's scalability.
    \item The LGB detection model comprises two primary modules: semantic understanding and structure extraction. In the semantic understanding module, we perform supervised fine-tuning on the LM using constructed user text sequences to enhance its semantic comprehension of the node's social information, thereby providing semantically enriched node representations for the entire system. In the structure extraction module, the GNN enhances the model's representation capabilities by extracting and integrating the structural information of social relationships into the semantically enriched node representations.
    \item We conduct extensive experiments on two public and independent datasets, and the results demonstrate the effectiveness of fusing social semantics with network structure to jointly detect accounts and the superior detection performance of LGB compared with various state-of-the-art baseline models. Furthermore, studies on online smart feedback and robustness prove the effectiveness of the online smart feedback function and the strong robustness of LGB.
\end{itemize}

\vspace{-0.4cm}
\vpara{Comparison with the conference version~\cite{zhou2023detecting} of this work, the following extensions are made:}
\vspace{-0.2cm}

\begin{itemize}[leftmargin=*,itemsep=0pt,parsep=0.5em,topsep=0.3em,partopsep=0.3em]
    \setlength{\itemsep}{0pt}
    \setlength{\parsep}{0pt}
    \setlength{\parskip}{0pt}
    \item We further analyze the social human-bot network data and find that up to $30.62\%$ of the total nodes are isolated nodes, about $24.71\%$ have only one neighbor, and only $8.2\%$ have more than 10 neighbors, as shown in Fig.~\ref{fig:Figure1_data_show}. These findings explain that the performance improvement bottleneck of traditional graph-based methods is the presence of a large number of isolated and sparsely linked nodes in the network. To address this issue, we explore the detection performance of LM and GNN for sparsely linked and densely linked nodes in Section~\ref{sec:lm_gnn_which_when_better} and find that the supervised fine-tuned LM can effectively detect sparsely linked nodes, while GNN is more effective for detecting densely linked nodes. These findings inspire us to combine LM and GNN to enhance the model's performance.
    \item Our structural analysis of social relationships in Section~\ref{sec:Structural_Analysis_of_Social_Relationships} reveals an intrinsic link between social relationship structure and bot probability. This finding underscores the importance of the structural information of social relationships for account detection tasks and motivates us to fuse node semantics with network structure for more effective social account detection.
    \item Inspired by the data analysis and findings in Section~\ref{sec:Human_bot_Network_Analysis}, we propose LGB in Section~\ref{sec:LGB}, a novel multimodal information fusion-driven social bot detection framework, to achieve efficient detection of both sparsely and densely linked nodes.
    \item To enhance the efficiency of multimodal information fusion-driven social bot detection, we design a new system architecture detailed in Section~\ref{sec:LGB}. Specifically, the LGB's offline training system is divided into two parts: LM and GNN model training, incorporating a multimodal information fusion operation. In addition, the data preprocessing module adds the unified user text sequence construction function.
    \item In Section~\ref{sec:Experiments}, a new large-scale dataset TwiBot-20~\cite{feng2021twibot} is added. Additionally, more performance comparison experiments, online smart feedback studies, ablation studies, and robustness studies are conducted to validate our model.
\end{itemize}

\vspace{-0.4cm}
\vpara{Organization:} The remaining sections of this paper are organized as follows. In Section~\ref{sec:Preliminary_and_Definition}, we formally define the social bot detection problem. In Section~\ref{sec:Human_bot_Network_Analysis}, we analyze the relationship between the structural diversity of social relationships and the probability of bots and answer the question of who is better and when, LM vs. GNN, through comparative experiments. In Section~\ref{sec:LGB}, we propose a novel LM and GNN-driven social bot detection framework, LGB, and introduce it in detail from both the system and model architecture levels. In Section~\ref{sec:Experiments}, extensive experimental results are shown. In Section~\ref{sec:related_work}, we present the related work, and finally we conclude this work in Section~\ref{sec:Conclusion}.

\vspace{-0.1cm}
\section{Preliminary and Definition}
\label{sec:Preliminary_and_Definition}

\subsection{Social Bot Detection}

Considering social users as nodes and social relationships as edges~\cite{zhou2023semi, zhang2024recdcl}, a social network can be regarded as a directed graph formally represented as $\mathcal{G}\left ( \mathcal{V},\mathcal{E} \right )$, where the set of edges $\mathcal{E}\subset \mathcal{V} \times \mathcal{V}$ represents social relationships between users, and the set of nodes $\mathcal{V} = \left\{ v_{1},v_{2}, ..., v_{N}\right\}$ represents social users. Let $ v_{i} $ indicate the node $i$ in graph $\mathcal{G}$, the set of its neighbors can be denoted as $ \mathcal{N} \left ( v_{i} \right ) = \left\{ u:\left ( v_{i}, u \right )\in \mathcal{E} \right  \} $. Let $\mathbf{h}_{i} \in \mathbb{R}^{d}$ represent the feature vector of node $i$, where $d$ denotes the feature dimension. The feature matrix of nodes in graph $\mathcal{G}$ can be represented as $\mathbf{H}=[\mathbf{h}_{1}$$,\mathbf{h}_{2}$$,\dots,\mathbf{h}_{N}]^\top$$\in$$\mathbb{R}^{N \times d}$. To formalize the social relationship between nodes, let the existence of an edge between node $i$ and node $j$ be represented as $1$, and the absence of an edge as $0$. The adjacency matrix $\mathbf{A} \in \mathbb{R}^{N \times N}$ of the graph $\mathcal{G}$ can be obtained, where $\mathbf{A}_{ij}$ denotes the relationship between node $i$ and node $j$. Based on $\mathbf{A}$, we can derive its diagonal degree matrix $\mathbf{D} \in \mathbb{R}^{N \times N} $, where $\mathbf{D}_{ij} = \sum_{j} \mathbf{A}_{ij}$ if $i = j$, otherwise $\mathbf{D}_{ij} = 0$.

\textit{Problem Formulation:} The purpose of this work is to identify whether a given social account $v_{i}\in \mathcal{V}$ is a human or a bot, which is treated as a node classification task. Specifically, the input is a social account $v_{i}$, and the system gathers its attribute information, personal description, and tweets to construct a unified user textual sequence $\mathbf{s}_{i} \in \mathcal{D}^{L_{i}}$, where $L_{i}$ denotes the length of its textual sequence, and $\mathcal{D}$ represents the dictionary of tokens or words. The detailed construction process of text sequences for social users will be presented in Section~\ref{sec:LGB}. The output is the predicted label $\widehat{y}_{i}=f\left ( \mathbf{s}_{i} \right )$ obtained by model inference, where $f\left ( \cdot  \right )$ denotes the model's inference function. Let $y_{i}\in \{0,1\}$ represent the ground truth of account $v_{i}$, where $y_{i} = 0$ means that $v_{i}$ is a normal user, while $y_{i} = 1$ means that $v_{i}$ is a social bot. Therefore, the goal of this study is to learn a function $f$ for $\widehat{y}_{i} \longrightarrow y_{i}$.

\vspace{-0.2cm}
\subsection{Language Models for Social Bot Detection}

 For the social bot detection task, LMs are used to extract the social semantic information from users' textual content and encode it into the feature matrix. Formally, let $\mathcal{LM}\left ( \cdot  \right )$ denote a text encoder based on a pre-trained language model, such as RoBERTa~\cite{liu2019roberta}, T5~\cite{raffel2020exploring}, etc. The textual sequence of node $i$ is represented as $\mathbf{s}_{i} \in \mathcal{D}^{L_{i}}$, and by feeding it into the pre-trained LM, we obtain the node representation, which is denoted as:
 \vspace{-0.4cm}

\begin{equation}
\label{eq:lm_hidden_state}
\vspace{-0.1cm}
\mathbf{x}_{i}=\frac{1}{\left| \mathbf{s}_{i}\right|}\sum_{t=1}^{\left|\mathbf{s}_{i} \right|}\mathcal{LM}\left ( \mathbf{s}_{i} \right )_{t},
\end{equation}
where $\left|\mathbf{s}_{i} \right|$ represents the number of tokens in the textual sequence $\mathbf{s}_{i}$. Here, we average the output of the LM for the token to obtain a representation vector $\mathbf{x}_{i}$, which is then fed into an MLP to predict the node category $\widehat{y}_{i}=$MLP$\left ( \mathbf{x}_{i} \right )$.

LMs, with their extensive semantic knowledge acquired from large-scale corpora in the pre-training stage and their significant number of parameters, have achieved success in numerous natural language tasks, such as text classification tasks. However, their huge model size results in high memory overhead. Moreover, in social networks, LMs only use each node's text information, neglecting social relationships and interaction information between nodes, which leads to their performance bottleneck, especially for nodes that lack text information.

\vspace{-0.2cm}
\subsection{Graph Neural Networks for Social Bot Detection}

Different from LMs, graph neural networks can aggregate the node representations of neighbors based on the social relationships between nodes to detect social accounts. To be specific, GNNs mainly consist of the following two steps: 1) message passing and aggregation, shown in \eqref{eq:gnn_aggregate}, and 2) updating node representations, shown in \eqref{eq:gnn_update}:

\vspace{-0.2cm}
\begin{equation}
\label{eq:gnn_aggregate}
\mathbf{m}_{i}^{\left ( l \right )} = \mathrm{AGGREGATE}^{\left ( l \right )}\left ( \left\{ \mathbf{h}_{j}^{\left ( l-1 \right )}:v_{j}\in \mathcal{N} \left ( v_{i} \right )\right\} \right ),
\end{equation}

\begin{equation}
\label{eq:gnn_update}
\mathbf{h}_{i}^{\left ( l \right )} = \mathrm{UPDATE}^{\left ( l \right )}\left (\mathbf{h}_{i}^{\left ( l-1 \right )} , \mathbf{m}_{i}^{\left ( l \right )}\right ),
\end{equation}
where $\mathbf{h}_{i}^{\left ( l \right )}$ represents the feature vector of node $i$ in layer $l$, and $\mathrm{UPDATE}^{\left ( l \right )}\left ( \cdot \right )$ is the update function of the $l$-th layer, which can be implemented by a neural network, such as an attention network or a multi-layer perceptron. By inputting the representation of node $i$ from the previous layer and the information aggregated from neighbors into this function, the next layer representation of node $i$ can be obtained. $\mathrm{AGGREGATE}^{\left ( l \right )}\left (  \cdot \right )$ is the aggregation function of the $l$-th layer, which usually adopts operations such as average, maximum, and summation. By inputting the node representations in the previous layer of the neighbors of node $i$ into this function, the aggregation vector $\mathbf{m}_{i}^{\left ( l \right )}$ of node $i$ in the $l$-th layer can be obtained. Next, the node representation containing network structure information is fed into a multi-layer perceptron with a softmax layer to identify the node category.

Based on the message-passing mechanism, graph neural networks~\cite{zhang2023apegnn, zhou2023detecting, zhang2023dropconn} can effectively utilize the structural and interactive information between nodes to detect social accounts, achieving superior results in social bot detection tasks. However, for isolated and sparsely linked nodes in social networks, GNNs suffer from performance degradation because of the lack of required social relationship information to enhance node representation.

The major notations used in this paper are listed in Appendix A. Before we start to analyze the human-bot network data, several key definitions to be used are described below.

\vspace{-0.2cm}
\vpara{Connected Components (CC)$\footnote{\url{https://en.wikipedia.org/wiki/Component_(graph_theory)}}$:} In graph theory, a component (also known as connected component) 
is a maximal connected subgraph of an undirected graph $\mathcal{G}$, and any two of its vertices are connected to each other. In this paper, we denote connected components as \textbf{CC} and the number of connected components as \textbf{NumCC}.

\vspace{-0.2cm}
\vpara{Ego network:} An ego network $\mathcal{G}\left ( \mathcal{V}_{v},\mathcal{E}_{v}  \right )$ is a subnetwork in a social network consisting of node $v$, named \textit{ego}, and its first-order neighbors, where $\mathcal{E}_{v}$ and $\mathcal{V}_{v}$ denote the edge set and node set in the ego network, respectively.

\vspace{-0.2cm}
\section{Human-bot Network Analysis}
\label{sec:Human_bot_Network_Analysis}
\vspace{-0.1cm}

To better motivate the core design of LGB, this section provides an empirical analysis of a popular social network, TwiBot-22~\cite{feng2022twibot}, which includes both humans and bots. Specifically, we investigate the comparative efficacy of LMs and GNNs for nodes with varying numbers of neighbors in Section~\ref{sec:lm_gnn_which_when_better}. Subsequently, we explore the structure of social relationships and examine their correlation with the distribution of account categories, thereby underscoring the significance of network structure in social bot detection, as detailed in Section~\ref{sec:Structural_Analysis_of_Social_Relationships}.

\vspace{-0.4cm}
\begin{figure}[htbp]
\setlength{\abovecaptionskip}{0.0cm}   
\setlength{\belowcaptionskip}{-0.3cm}   
\centering
\includegraphics[width=0.5\linewidth]{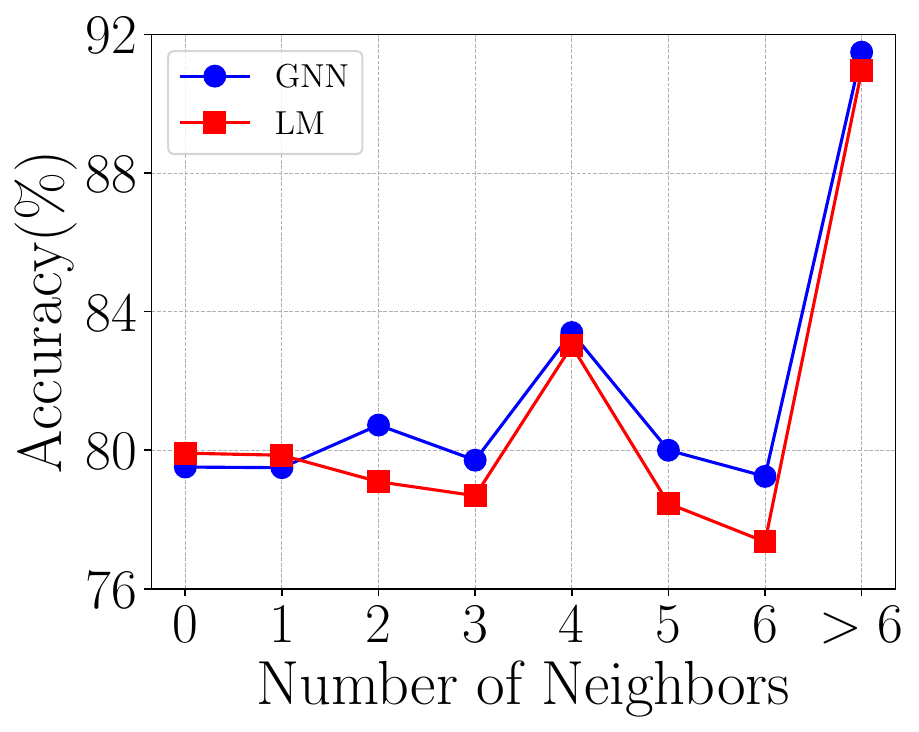}
\caption{{LM vs. GNN for nodes with different numbers of neighbors. X-axis: the number of neighbors; Y-axis: the detection accuracy of models.}} \label{fig:lm_vs_gnn_acc}
\vspace{-0.6cm}
\end{figure}

\vspace{-0.2cm}
\subsection{Comparative Analysis of LMs and GNNs}
\label{sec:lm_gnn_which_when_better}

By analyzing the neighbor distribution of nodes in the social network depicted in Fig.~\ref{fig:Figure1_data_show}, we observe that up to 30.62\% of the nodes are isolated, and approximately 24.71\% of the nodes have only one neighbor. In contrast, nodes with more than ten neighbors account for only 8.2\%. This indicates a significant presence of isolated nodes and nodes with few links in the social network. For graph-based methods, isolated nodes can cause the aggregation vector $\mathbf{m}_{i}^{\left ( l \right )} = 0$ in \eqref{eq:gnn_aggregate} and \eqref{eq:gnn_update}, which degenerates the GNN into a multi-layer perceptron and weakens its detection performance.

In such a case, a powerful representation model for node features becomes a more promising choice. Motivated by the strong representation modeling capability of LMs~\cite{du2022glm,zhang2024sciglm,lilong2024autore,zhang2024rest}, we perform supervised fine-tuning to align them to the social bot detection task, and experimentally compare the detection accuracy of LMs and GNNs for nodes with different numbers of neighbors. Specifically, RoBERTa~\cite{liu2019roberta} and GIN~\cite{xu2018powerful} are chosen for LMs and GNNs in our experiments, respectively. The results are plotted in Fig.~\ref{fig:lm_vs_gnn_acc}.

The comparison reveals that for isolated nodes and nodes with only one neighbor, the LM achieves higher detection accuracy. In contrast, for nodes with more than two neighbors, the GNN performs better. This can be attributed to the fact that for isolated nodes and nodes with few neighbors, the advantage of GNNs in modeling graph structure information diminishes due to the lack of social relationships. Meanwhile, LMs, having absorbed extensive social semantic knowledge during pre-training, can transfer this knowledge effectively through supervised fine-tuning, thereby producing informative representations for isolated and sparsely linked nodes, which benefits bot detection. As the number of neighbors increases, GNNs can effectively leverage social relationships to aggregate information from neighbors to the central node, thereby enhancing node representation and improving detection accuracy. 

For nodes with more than six neighbors or exactly four neighbors, as shown in Fig.~\ref{fig:lm_vs_gnn_acc}, the presence of more edges may introduce noise from neighbors, causing some performance fluctuations. However, the overall trend shows that as the number of neighbors increases, GNNs consistently outperform LMs in detection accuracy. The importance of structural information in social relationships will be further examined in Section~\ref{sec:Structural_Analysis_of_Social_Relationships}.

Based on the above analysis, we conclude that for isolated nodes and nodes with few neighbors, LMs can achieve more effective detection by understanding the semantic information of social accounts; as the number of neighbor nodes increases, GNNs can achieve higher detection performance by capturing network structure information. These findings suggest that combining LMs and GNNs to exploit both social semantics and network structure can enhance detection performance.

\vspace{-0.3cm}
\subsection{Structural Analysis of Social Relationships}
\label{sec:Structural_Analysis_of_Social_Relationships}

\begin{figure*}[t!]
\setlength{\belowcaptionskip}{-0.8cm}   
\centering
\subfloat[]{\includegraphics[width=0.32\linewidth]{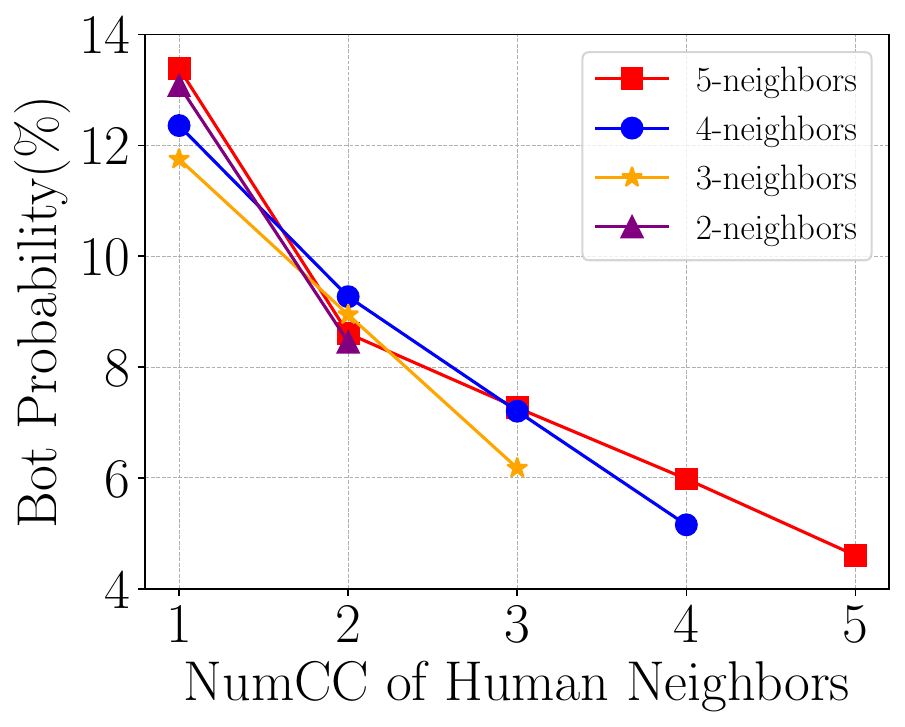}%
\label{subfig:tw22_connected_component_analysis_human}}
\hfil
\centering
\subfloat[]{\includegraphics[width=0.32\linewidth]{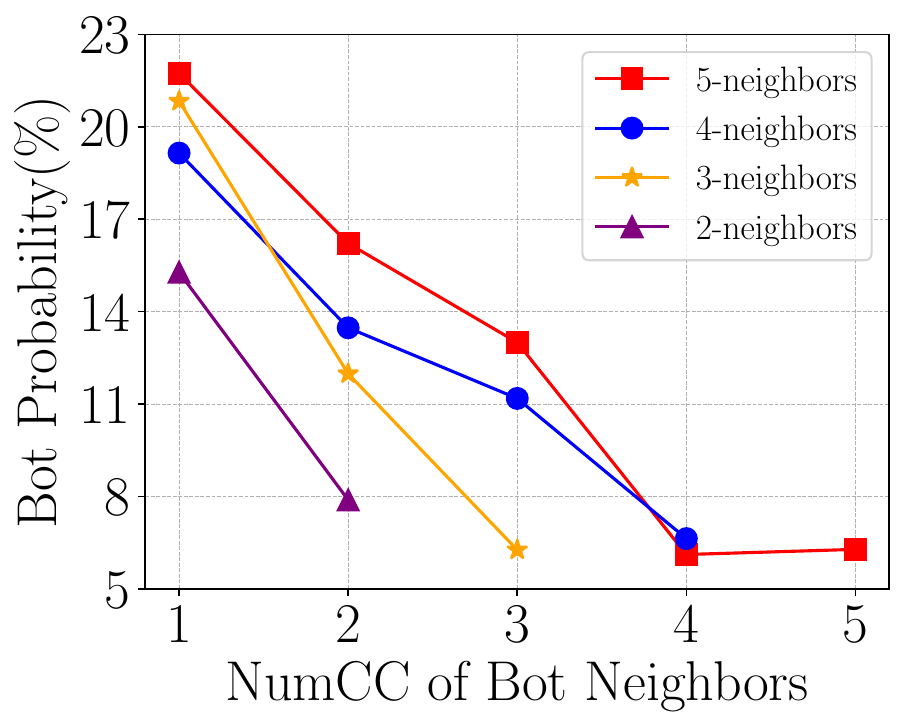}%
\label{subfig:tw22_connected_component_analysis_bot}}
\hfil
\centering
\subfloat[]{\includegraphics[width=0.32\linewidth]{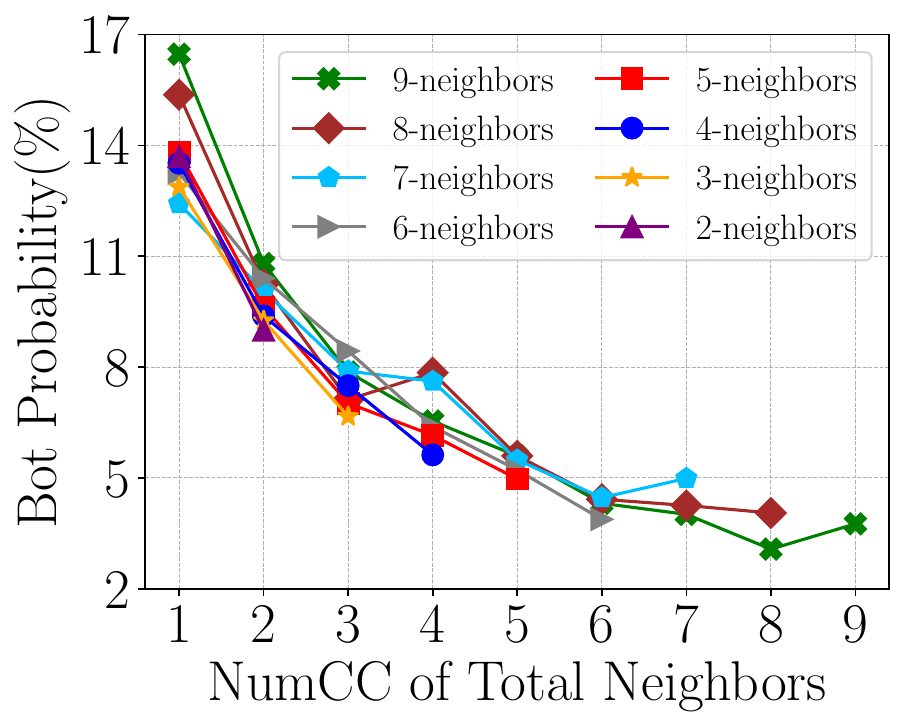}%
\label{subfig:tw22_connected_component_analysis}}
\caption{Social relationship structure analysis. The Y-axis represents bot probability, and the X-axis indicates the number of connected components (NumCC) of (a) human neighbors, (b) bot neighbors, or (c) total neighbors.}
\label{fig:social_structure_analysis}
\vspace{-0.5cm}
\end{figure*}

In Section~\ref{sec:lm_gnn_which_when_better}, we have analyzed the impact of the number of neighbors (i.e., different numbers of edges) on detection performance. Inspired by structural diversity~\cite{ugander2012structural, zhang2013social}, which suggests that different social relationship structures affect users' behavior differently, we explore the relationship between the structure of a user's friend circle and its bot probability below. Specifically, we first export users' ego networks, which are the induced subgraphs formed by their friend circles. Next, we count the number of connected components (NumCC) of human neighbors (Fig.~\ref{subfig:tw22_connected_component_analysis_human}), bot neighbors (Fig.~\ref{subfig:tw22_connected_component_analysis_bot}), and total neighbors (Fig.~\ref{subfig:tw22_connected_component_analysis}) in ego networks with varying numbers of friends, and analyze users' bot probability under different NumCC. Our analysis yields several intriguing discoveries:

\vspace{-0.2cm}
\vpara{CC Analysis for Human Neighbors.} Fig.~\ref{subfig:tw22_connected_component_analysis_human} shows that the bot probability decreases as the number of connected components (NumCC) of human neighbors increases. In social activities, social bots influence humans' behavior through the spread of misinformation, whose scope is determined by the time and speed of transmission. A network structure with fewer connected components is more cohesive, leading to faster propagation of information. For propagation time, social researchers find that more structural diversity makes more knowledge sharing~\cite{cummings2004work}, allowing people to quickly verify the correctness of messages. Therefore, in the ego networks of human neighbors, more connected components are not good for the spread of misinformation and bots' survival.

\vspace{-0.2cm}
\vpara{CC Analysis for Bot Neighbors.} Fig.~\ref{subfig:tw22_connected_component_analysis_bot} illustrates that the bot probability decreases as the number of connected components (NumCC) of bot neighbors increases. Studies find that bots collaborate to spread misinformation and carry out malicious behavior, reducing the risk of being detected~\cite{zhang2016rise}. Fewer connected components make the bots' networks more cohesive, facilitating bot collaboration. So, in the ego networks of bot neighbors, more connected components are not good for the interaction of bots and their gang sabotage.

\vspace{-0.2cm}
\vpara{CC Analysis for Total Neighbors.} Fig.~\ref{subfig:tw22_connected_component_analysis} shows that the bot probability decreases as the number of connected components (NumCC) of total neighbors increases. This is the superposition of ego networks of human neighbors and bot neighbors, indicating that less structural diversity facilitates the spread of misinformation and bot collaboration.

The above structural analysis of social relationships reveals an inherent link between bot probability and social relationship structure, underscoring the significance of social relationship structure in social account detection tasks. This finding motivates us to integrate social relationship structure with account semantics to improve the detector's performance.

\vspace{-0.4cm}
\subsection{Summaries} \label{Summaries}
We get the following discoveries from the above analyses:

\vspace{-0.2cm}
\begin{itemize}[leftmargin=*,itemsep=0pt,parsep=0.5em,topsep=0.3em,partopsep=0.3em]
    \setlength{\itemsep}{0pt}
    \setlength{\parsep}{0pt}
    \setlength{\parskip}{0pt}
    \item Aligning LMs to the social account detection task through supervised fine-tuning enables them to fully exploit the textual semantics of accounts, thereby achieving effective detection of isolated and less-linked nodes compared to graph-based methods. While, as the number of edges connected to nodes (i.e., the number of neighbors) increases, graph-based methods can achieve more effective account detection than LMs by capturing network structure information.
    \item By analyzing the relationship between the structure of users' social relationships and the probability that they are social bots, we find that the probability of social accounts being bots is negatively correlated to the number of connected components formed by humans or bots in their friend circle.
\end{itemize}

\begin{figure*}[t!]
\setlength{\abovecaptionskip}{0.0cm}   
\centering
\includegraphics[width=0.96\linewidth]{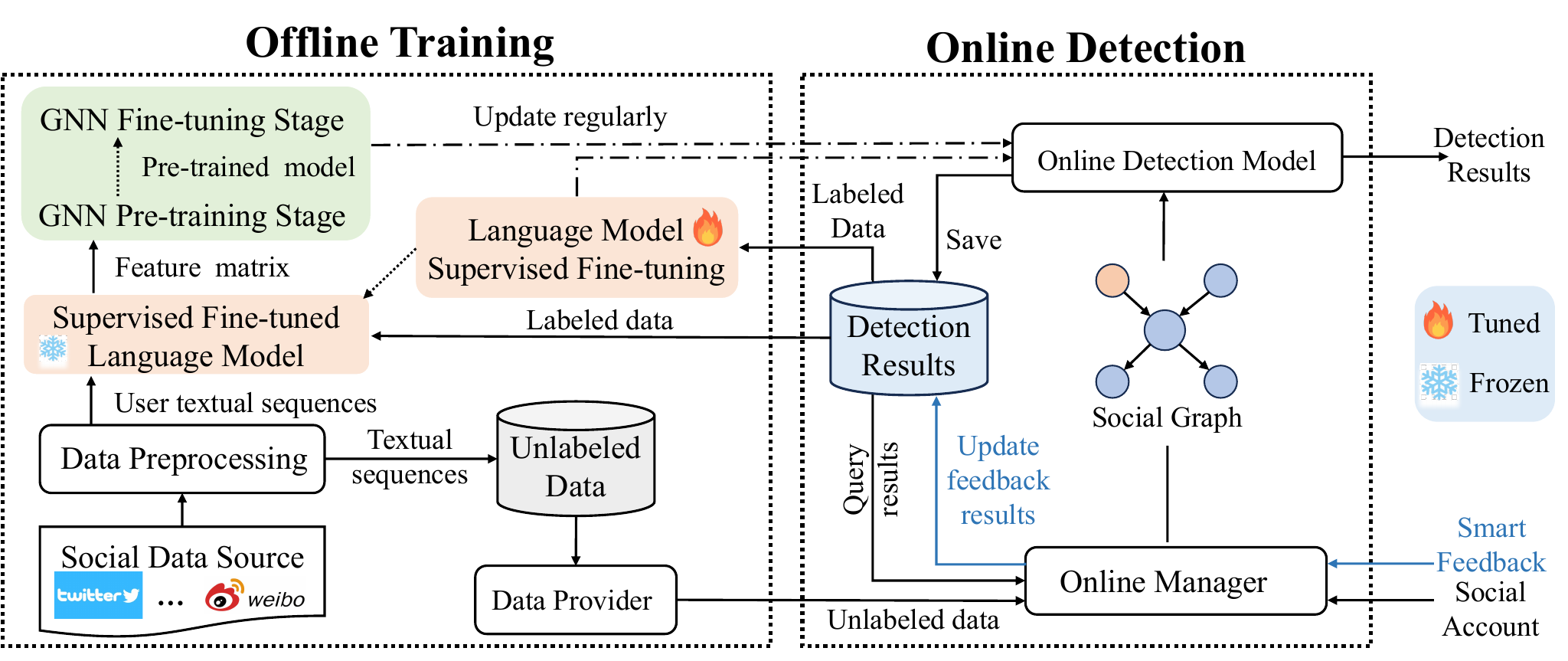}
\caption{{The overall system architecture of LGB primarily comprises two subsystems: offline training and online detection. These subsystems work collaboratively through data interaction.}} 
\label{fig:overall_framework}
\vspace{-0.5cm}
\end{figure*}

\vspace{-0.4cm}
\section{LGB Framework}
\label{sec:LGB}

Through the analysis of social network data, as illustrated in Fig.~\ref{fig:Figure1_data_show}, we have identified numerous isolated and sparsely linked nodes that can weaken graph-based approaches. To address this issue, inspired by the comparative experiments and data analysis in Section~\ref{sec:Human_bot_Network_Analysis}, we propose a \textbf{L}anguage model and \textbf{G}raph neural network-driven social \textbf{B}ot detection framework (\textbf{LGB}). This framework jointly utilizes LMs and GNNs to capture bimodal information of node semantics and network structure, achieving high-performance social bot detection.

\vspace{-0.3cm}
\subsection{Framework Overview}

The overall system architecture of LGB is depicted in Fig.~\ref{fig:overall_framework}. It consists of two subsystems: offline model training and online real-time detection. These subsystems collaborate through continuous data interaction to enable real-time social account detection with a smart feedback function while enhancing the system's scalability to meet the needs of distributed deployment. The working principles of the system are as follows:

\vspace{-0.2cm}
\vpara{Offline Training.} This part provides model offline training and data preprocessing services. As highlighted in the offline training subsystem of Fig.~\ref{fig:overall_framework}, model training primarily involves the optimization of LMs and GNNs, allowing for the flexible deployment of various models and training strategies. Data preprocessing services include data collection, processing, and persistence, with the data source collected from major social platforms such as Weibo and Twitter/X.
Additionally, LGB offers data acquisition and storage tools and information processing components, all supporting distributed deployment.

\vspace{-0.2cm}
\vpara{Online Detection.} This part is mainly responsible for online real-time account detection and processing user feedback information. Specifically, logged-in users can submit feedback to the system for questionable account detection results, which, if validated, is updated in the detection results database. The updated data is then provided to the offline training subsystem for the next round of model training. Subsequently, the model, which is regularly trained offline, is sent back to the online detection subsystem to update its detection model, ensuring the latest knowledge is applied to online account detection. Besides the account detection functionality on the web page, APIs supporting batch detection are provided, allowing social applications to incorporate malicious account identification capabilities for safer online socialization.

In the following sections, data preprocessing and model training of the offline subsystem are detailed in Section~\ref{sec:offline_sys_data_service} and Section~\ref{sec:offline_sys_training}, respectively. Section~\ref{sec:Online System} then introduces the principles of real-time online detection and smart feedback.

\vspace{-0.4cm}
\subsection{Data Preprocessing}
\label{sec:offline_sys_data_service}

The data preprocessing module is designed to perform essential functions such as data collection, processing, and storage, thereby constructing a comprehensive social information database.

\subsubsection{Graph Collection} The process of social graph collection comprises three key stages: seed user selection, graph expansion, and feature alignment.

\textit{Seed User Selection.} The seed user selection phase establishes a foundational pool of influential social users spanning ten social domains. From this pool, a subset of seed users is chosen to form a seed set $S$. \textit{Graph Expansion.} Based on the selected seed users, the breadth-first search method is employed to expand the social graph. Initially, seed users $s_{i}\in S$ are integrated into the graph as unique nodes during the first iteration. Subsequently, at each iteration, the followers and followed users of each node are incorporated into the graph, along with their corresponding follow relationships. \textit{Feature Alignment.} During this stage, comprehensive social information, including account attributes, tweets, comments, likes, and reposts, is gathered for each node within the constructed social graph.

\subsubsection{Construction of Unified User Textual Sequences}

\begin{figure}[htbp]
\vspace{-0.3cm}
\setlength{\abovecaptionskip}{0.0cm}   
\centering
\includegraphics[width=\linewidth]{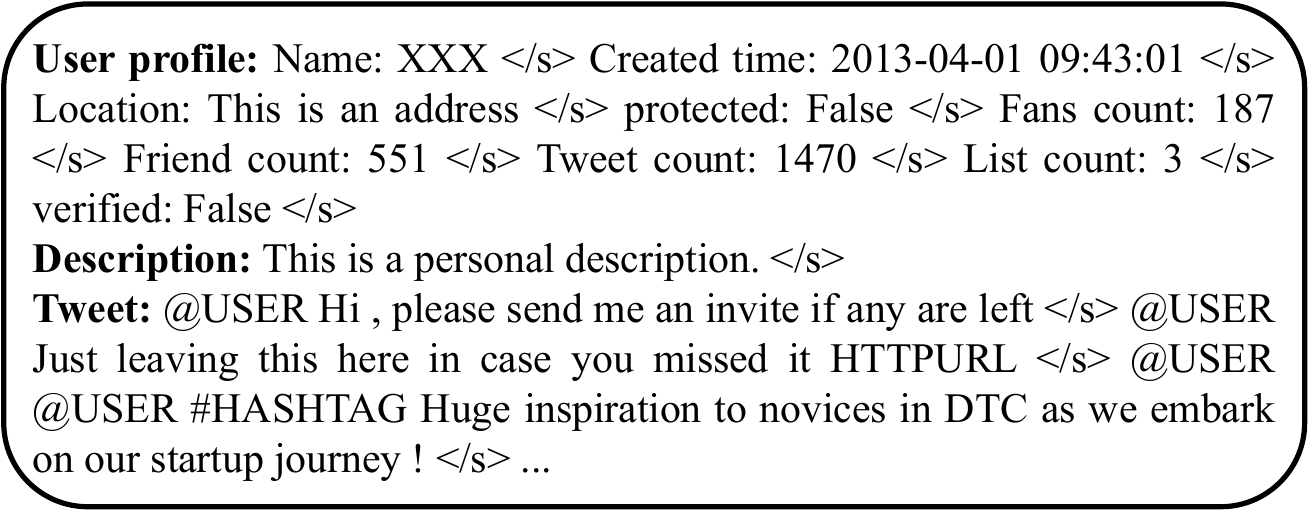}
\caption{{The unified user textual sequence.}} \label{fig:user_text_sequence}
\vspace{-0.3cm}
\end{figure}

During the phase of social graph collection and construction, various user attributes, personal descriptions, and tweets are collected and stored. To conform to the text input requirements of LMs, we create unified textual sequences for users. Initially, user attribute information (such as name, fans count, and friend count), personal descriptions, and tweets are extracted from the raw data and separately organized. These components are then merged into cohesive sequences, as illustrated in Fig.~\ref{fig:user_text_sequence}, where \textbf{User profile}, \textbf{Description}, and \textbf{Tweet} represent the initial symbols denoting user attributes, personal descriptions, and tweets, respectively. The delimiter $\displaystyle</s>$ indicates the boundary between segments.

Given that social user information often contains noise such as emoticons, mentions, hashtags, and web addresses, which may hinder LMs' comprehension of textual content, noise reduction techniques are applied. Initially, TweetTokenizer\cite{TweetTokenizer_web} is employed to tokenize user text information. Subsequently, emoticons, mentions, hashtags, and web addresses are replaced with text descriptions of emoticons, @USER, \#HASHTAG, and HTTPURL, respectively. This process yields unified textual sequences, which are then fed into supervised fine-tuned LMs for encoding to generate user feature matrices. These textual sequences are stored in the offline system's database for subsequent model training and account detection. Each data query is routed to the data provider, and if the requested data record is absent in the database, real-time data collection and processing are initiated.

\begin{figure*}[t!]
\setlength{\abovecaptionskip}{0.0cm}   
\centering
\includegraphics[width=\linewidth]{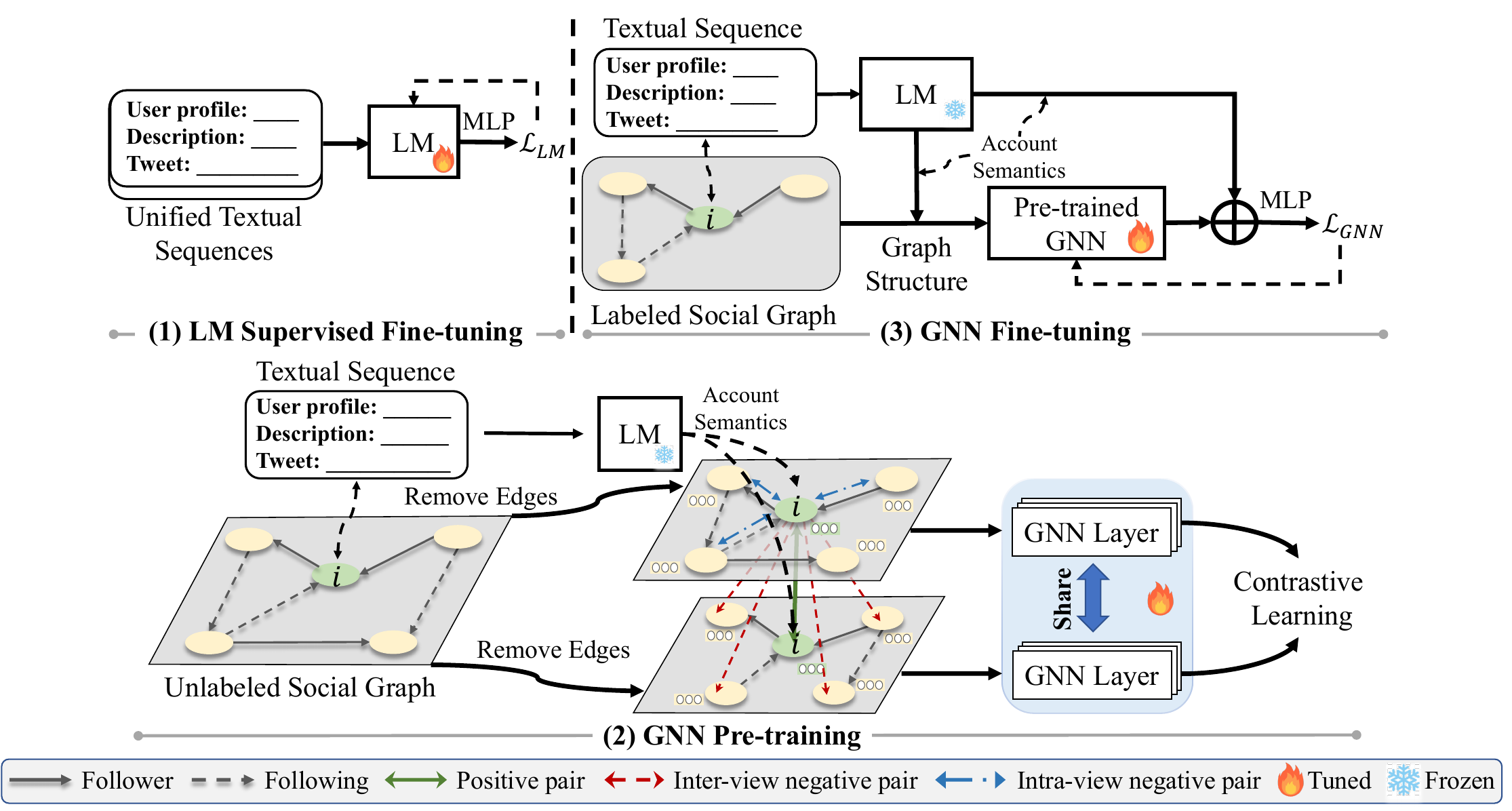}
\caption{{Illustration of the model architecture of LGB: (1) The LM is fine-tuned with supervision based on online annotated data to better align with the social bot detection task, thereby enhancing its understanding of social semantics. (2) The GNN acquires valuable knowledge from unlabeled data through graph contrastive learning. (3) The GNN is fine-tuned to improve the overall detection performance by integrating the bimodal information of account semantics and network structure.}} \label{fig:lgb_model}
\vspace{-0.5cm}
\end{figure*}

\vspace{-0.2cm}
\subsection{Model Learning}
\label{sec:offline_sys_training}

The goal of offline model training is to extract semantic information from social accounts and structural information from social relationships using both collected unlabeled data and online labeled data. This process aims to enhance the detection of social accounts. As illustrated in the upper left part of Fig.~\ref{fig:overall_framework}, offline model training is divided into two parts: LM and GNN.

In the LM part, supervised fine-tuning is employed on online annotated data to align the LM to the account detection task. The GNN part is bifurcated into two stages: pre-training with unlabeled data and fine-tuning with online labeled data. This multimodal staged offline training architecture enables the model to effectively mine semantic information from nodes and structural information from social networks, leveraging both large-scale unlabeled data and continuously growing online annotated data. Consequently, the model achieves efficient detection of both sparsely and densely linked nodes. The subsequent sections detail the LM and GNN parts of LGB.

\vspace{-0.2cm}
\vpara{Supervised fine-tuning of language models.} Inspired by experimental results that demonstrate the effectiveness of supervised fine-tuning of LMs in detecting isolated and less-linked nodes (Section~\ref{sec:lm_gnn_which_when_better}), the offline subsystem initiates each training round using annotated data derived from online user feedback and detection results (detailed in Section~\ref{sec:Online System}). This data, enriched with network structure knowledge from the GNN and user feedback, is utilized for supervised fine-tuning of the LM, aligning it with the social account detection task. The integration of network structure knowledge and user feedback into the LM is depicted in the upper left of Fig.~\ref{fig:lgb_model}. Specifically, unified textual sequences $\mathbf{S}$ are processed by a language model $\mathcal{LM}$ to generate node representation matrix $\mathbf{X}$, which is then fed into a multi-layer perceptron with softmax to produce prediction results $\hat{\mathbf{Y}}$:
\vspace{-0.1cm}
\begin{equation}
\label{eq:lm_prediction_y}
\hat{\mathbf{Y}} = \mathrm{softmax}\left ( \mathrm{MLP}\left ( \mathbf{X} \right ) \right ),\quad \mathbf{X} = \mathcal{LM}\left ( \mathbf{S} \right ).
\end{equation}

\vspace{-0.1cm}
The LM is optimized and aligned to the social account detection task with the following objective:
\vspace{-0.2cm}
\begin{equation}
\label{eq:loss_lm}
\mathcal{L}_{LM} = \frac{1}{|\mathbf{\Omega}_{l}|}\sum_{i \in \mathbf{\Omega}_{l}} \mathrm{CrossEntropy}\left(\mathbf{Y}_{i}, \widehat{\mathbf{Y}}_{i} \right),
\vspace{-0.2cm}
\end{equation}
where $\mathbf{Y}_{i}$ and $\widehat{\mathbf{Y}}_{i}$ represent the ground truth and prediction result for node $i$, respectively. $\mathbf{\Omega}_{l}$ denotes the annotated data incorporating network structure knowledge learned by GNN and user feedback.

\vspace{-0.2cm}
\vpara{GNN pre-training based on GCL.} Building on the success of CBD~\cite{zhou2023detecting} in addressing the scarcity of annotated data in social bot detection through graph contrastive learning (GCL), a similar approach is employed in the GNN pre-training stage of LGB. This method leverages GCL to extract valuable insights from newly collected unlabeled data, thereby enhancing the model's capacity to capture structural information within social networks, as illustrated in the lower part of Fig.~\ref{fig:lgb_model}.
Initially, the supervised fine-tuned LM encodes unified user textual sequences to produce the user feature matrix $\mathbf{X}$. This matrix, combined with the adjacency matrix $\mathbf{A}$ representing the network structure, forms a social graph $\mathcal{G} = \left ( \mathbf{X}, \mathbf{A} \right )$ as the GNN model input. By randomly removing edges from $\mathbf{A}$, two views $\widetilde{\mathcal{G}}_{1} = \left ( \mathbf{X}, \widetilde{\mathbf{A}}_{1} \right )$ and $\widetilde{\mathcal{G}}_{2} = \left ( \mathbf{X}, \widetilde{\mathbf{A}}_{2} \right )$ are generated from the graph $\mathcal{G}$. These views are fed into the GNN to obtain the feature matrices incorporating network structure information:
\begin{equation}
\label{eq:h_m}
\mathbf{H}_{(m)} = \mathrm{GNN}\left ( \mathbf{X}, \widetilde{\mathbf{A}}_{m} \right ) \in \mathbb{R}^{N \times d},
\end{equation}
where $\widetilde{\mathbf{A}}_{m}$ and $\mathbf{H}_{(m)}$ denote the adjacency matrix and node representation of $\widetilde{\mathcal{G}}_{m} (m=1,2)$, respectively. $\mathrm{GNN}(\cdot)$ represents the GNN encoder.

For the representations $\mathbf{H}_{(1)}$ and $\mathbf{H}_{(2)}$ of the two views, the goal of contrastive learning is to maximize the distinction between representations of the same nodes and other nodes. Specifically, the representations $\mathbf{H}_{i, (1)}$ and $\mathbf{H}_{i, (2)}$ of the same node $i$ form a positive pair, while representations of different nodes form negative pairs. The InfoNCE~\cite{oord2018representation} loss for any node $i$'s positive pair is computed as follows:
\begin{equation}
\label{eq:contrastive_obj}
\mathcal{L}\left (\mathbf{H}_{i,(1)}, \mathbf{H}_{i,(2)} \right )= -\log\frac{e^{\mathrm{Sim}\left (\mathbf{H}_{i,(1)}, \mathbf{H}_{i,(2)} \right )/\tau}}{e^{\mathrm{Sim}\left (\mathbf{H}_{i,(1)}, \mathbf{H}_{i,(2)} \right )/\tau} + Neg},
\end{equation}
where $\tau$ is the temperature hyperparameter, and $\mathrm{Sim}(\cdot, \cdot)$ denotes the similarity function (e.g., cosine similarity). The term $\mathrm{Neg}$ represents the penalty from negative pairs:
\begin{equation}
\label{eq:neg}
Neg = \sum_{i \neq j} e^{\mathrm{Sim}\left (\mathbf{H}_{i,(1)}, \mathbf{H}_{j,(2)} \right )/\tau} + e^{\mathrm{Sim}\left (\mathbf{H}_{i,(1)}, \mathbf{H}_{j,(1)} \right )/\tau},
\end{equation}
where the first part penalizes inter-view negative pairs, and the second part penalizes intra-view negative pairs.

Given the symmetry of the two views, their loss functions are similar. The total loss for the graph pre-training stage is:
\begin{equation}
\label{eq:overall_cl}
\mathcal{L}_{GCL} = \frac{1}{2|\mathbf{\Omega}_{u}|} \sum_{i \in \mathbf{\Omega}_{u}} \left [ \mathcal{L}\left (\mathbf{H}_{i,(1)}, \mathbf{H}_{i,(2)} \right ) + \mathcal{L}\left (\mathbf{H}_{i,(2)}, \mathbf{H}_{i,(1)} \right ) \right ],
\end{equation}
where $\mathbf{\Omega}_{u}$ indicates the unlabeled data collected in the offline subsystem.

\vspace{-0.2cm}
\vpara{GNN fine-tuning based on multi-modal fusion.} The pre-trained GNN model is fine-tuned using annotated data derived from online detection results and user feedback, which will be detailed in Section~\ref{sec:Online System}. This fine-tuning process aims to further align the model with the social bot detection task. To effectively detect isolated and sparsely linked nodes in social networks, we utilize a multi-modal fusion approach at this stage, integrating semantic knowledge learned by the LM into the GNN. As depicted in the upper right part of Fig. \ref{fig:lgb_model}, the annotated data based on online detection results and user feedback is fed into both the supervised fine-tuned LM and the pre-trained GNN model to extract semantic and graph structure information, respectively. This process is represented as follows:
\vspace{-0.1cm}
\begin{equation}
\label{eq:ft_gnn}
\mathbf{H} = \mathrm{GNN}\left(\mathbf{X}, \mathbf{A}\right);\quad \mathbf{X} = \mathcal{LM}\left(\mathbf{S}\right),
\end{equation}
\vspace{-0.1cm}
where $\mathbf{S}$ represents the users' textual sequences from the online detection results database. These sequences are encoded by the LM to produce the users' feature matrix $\mathbf{X}$. By inputting the adjacency matrix $\mathbf{A}$ and the feature matrix $\mathbf{X}$ into the GNN model, we obtain the node representation $\mathbf{H}$, which incorporates network structure information.

To further enhance the model's representation capability by fusing node semantics and social relationship structure, we concatenate the outputs of the LM and GNN. This combined output is then processed through an MLP for information fusion, leading to the final predicted result $\hat{\mathbf{Y}}$ for the nodes, obtained via the softmax function:
\begin{equation}
\label{eq:ft_concat}
\hat{\mathbf{Y}} = \mathrm{softmax}\left(\mathrm{MLP}\left(\mathrm{Concat}\left(\mathbf{X}, \mathbf{H}\right)\right)\right),
\end{equation}
where $\mathrm{Concat}\left(\cdot, \cdot\right)$ is the concatenation function. The optimization goal during the GNN fine-tuning stage is to minimize the cross-entropy loss between the predicted result $\widehat{\mathbf{Y}}_{i}$ and the ground truth $\mathbf{Y}_{i}$ for each node, which is formulated as follows:
\begin{equation}
\label{eq:loss_gnn}
\mathcal{L}_{GNN} = \frac{1}{|\mathbf{\Omega}_{l}|} \sum_{i \in \mathbf{\Omega}_{l}} \mathrm{CrossEntropy}\left(\mathbf{Y}_{i}, \widehat{\mathbf{Y}}_{i}\right),
\end{equation}
where $\mathbf{\Omega}_{l}$ represents annotated data based on online detection results and user feedback, encompassing node semantic knowledge learned by the LM and user feedback information.

Through the above offline model training, we develop a model that integrates account semantics and network structure multi-modal information. This model is regularly deployed to the online system for ongoing account detection, as discussed in Section~\ref{sec:Online System} below.

\vspace{-0.3cm}
\subsection{Online Detection and Smart Feedback}
\label{sec:Online System}

The online detection subsystem primarily consists of the online manager and the detection model (shown in the right half of Fig.~\ref{fig:overall_framework}), offering two key services: \textit{real-time social account detection} and \textit{smart feedback}. The principles and workflows of these services are elaborated below.

\begin{figure}[t!]
\centering
\setlength{\abovecaptionskip}{0.0cm}   
\includegraphics[width=0.4\textwidth]{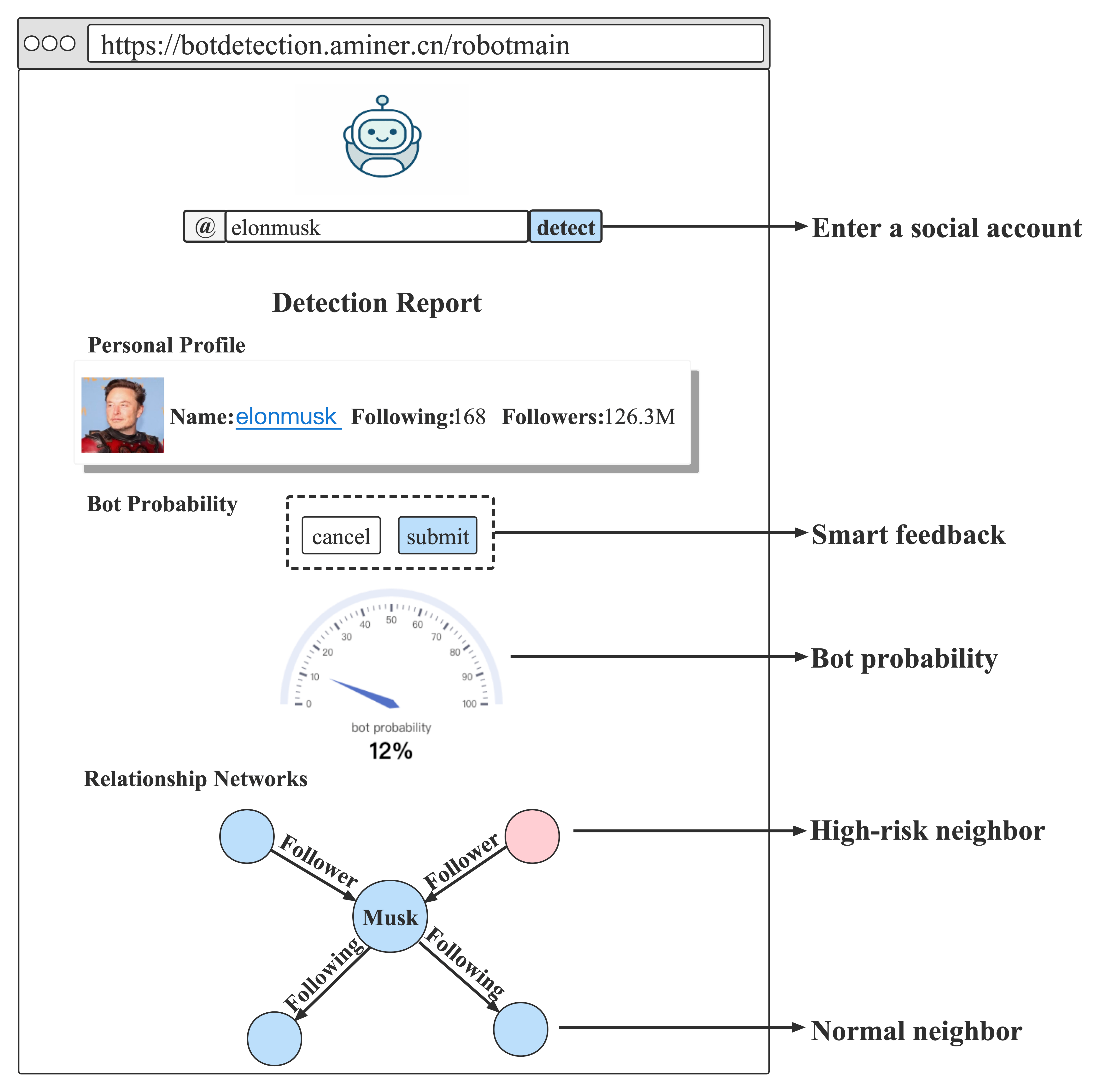}
\caption{Functional demonstration of LGB. The personal profile module displays the basic information of the detected account. The bot probability module indicates the likelihood that this user is a social bot. The relationship network illustrates the detection results for the account and its neighbors. High-risk bot accounts are marked in red, while other accounts are marked in blue.} \label{fig:system_website_show}
\vspace{-0.6cm}
\end{figure}

\vspace{-0.2cm}
\vpara{Online real-time social account detection.} For accounts suspected of being bots in social networks, users can input them into the detection box and click the detection button on the detection system website (accessible at \url{https://botdetection.aminer.cn/robotmain}) to initiate detection, as depicted in Fig.~\ref{fig:system_website_show}. When the online manager receives an account detection request, it first checks if the detection result for that account already exists in the detection results database. If the result is not found, the manager requests relevant information about the account from the data provider and forwards it to the online detection model for real-time analysis. Specifically, the ego network $\mathcal{G}$ of the detected target node is constructed in real-time and analyzed by the online detection model. Based on the social network $\mathcal{G}$, the model assesses the target node and its neighbors, ultimately generating a risk detection report for the entire network, which is displayed in the lower part of Fig. \ref{fig:system_website_show}. In addition, these detection results are stored in the detection results database for use in subsequent rounds of offline model training. This approach allows the online bimodal information of account semantics and network structure to be continuously incorporated into the offline model training process, thereby enhancing its performance.

\vspace{-0.2cm}
\vpara{Smart feedback.} Recent research~\cite{cresci2020decade} has revealed that social bots are rapidly evolving to evade detection. To counter this, we have introduced an online smart feedback function, as shown in the middle part of Fig.~\ref{fig:system_website_show}. This function enables the model to continuously acquire the latest bot information provided by expert users, facilitating quick self-upgrades for effective detection of new bots. Specifically, when expert users question the model's detection results, they can submit feedback to correct them. This feedback undergoes a review process by both machines and humans. If approved, the feedback is recorded in the detection results database and incorporated into the next round of offline training, allowing the model to learn knowledge from human experts. The effectiveness of this smart feedback mechanism will be evaluated through an online smart feedback study in Section~\ref{section:Online_Smart_feedback_Study}.

\vspace{-0.1cm}
\section{Experiments}
\label{sec:Experiments}

\subsection{Experimental Setup}

\vpara{Datasets.} To verify the performance of LGB on real social networks, we use two independent and publicly available datasets collected from real social networks, namely TwiBot-22~\cite{feng2022twibot} and TwiBot-20~\cite{feng2021twibot}. Their statistical information is shown in Table~\ref{Dataset_statistics}. 
We randomly select $81,432$ social bots and $81,433$ normal users from TwiBot-22, constructing a sampling set containing $162,865$ accounts in total. This set is randomly divided into training, validation, and test sets in a ratio of $7:2:1$ to ensure the fairness of the experiments. For TwiBot-20, we adopt the same data settings as in \cite{feng2021twibot}.

\begin{table}[htbp]
\vspace{-0.3cm}
\caption{Dataset statistics.} 
\large
\begin{center}
\scalebox{0.70}{
\begin{tabular}{@{}ccccccc@{}}
\toprule
\textbf{Dataset}                       & \textbf{Human} & \textbf{Bot} & \textbf{Nodes}  & \textbf{Edges}  & \textbf{Classes}  \\ \midrule
TwiBot-22~\cite{feng2022twibot}        & 81,433         & 81,432       & 162,865         & 151,841         & 2                 \\ \midrule
TwiBot-20~\cite{feng2021twibot}        &  5,237         & 6,589        & 229,580         & 33,716,171      & 2                 \\ \bottomrule
\end{tabular}
}
\end{center}
\label{Dataset_statistics}
\vspace{-0.5cm}
\end{table}

\vpara{Baselines.} In the comparative experiments, we use three representative GNN models, namely, GIN~\cite{xu2018powerful}, GCN~\cite{kipf2016semi}, GAT~\cite{velickovic2017graph}, and the general language model RoBERTa~\cite{liu2019roberta} to build the LGB model, and we compare them with twelve baseline models. These baseline models include two general GNN models: GCN~\cite{kipf2016semi} and GIN~\cite{xu2018powerful}; six advanced GNN models: GCNII~\cite{chen2020simple}, GPR-GNN~\cite{chien2020adaptive}, MixHop~\cite{abu2019mixhop}, APPNP~\cite{klicpera2018predict}, LINKX~\cite{lim2021large}, and H2GCN~\cite{zhu2020beyond}; two recently released state-of-the-art social bot detection models: SIRAN~\cite{zhou2023semi} and CBD~\cite{zhou2023detecting}; two popular large-scale language models (LLMs): Vicuna-7B-v1.5~\cite{zheng2023judging} and ChatGLM3-6B~\cite{du2022glm,zeng2022glm}. The same unified user textual sequences, as shown in Fig.~\ref{fig:user_text_sequence}, are used as input for all experiments. More details of the baselines are described in Appendix B.

\vspace{-0.1cm}
\vpara{Implementation details.} Based on the directional attributes of social relationships between users, we construct the social network data as a directed graph, where the content of each node is the user's social information described by the unified user textual sequence, which is shown in Fig.~\ref{fig:user_text_sequence}. The AdamW optimizer~\cite{kingma2014adam} is employed during model training and optimization. For the LM part of the LGB model, weight decay and learning rate in the supervised fine-tuning phase are $10^{-2}$ and $10^{-5}$ respectively. For the GNN part of the LGB model, weight decay is $10^{-5}$, and the learning rate is set differently in different training stages. Specifically, in the pre-training stage, the learning rate is $10^{-3}$ on both TwiBot-22 and TwiBot-20, and in the fine-tuning stage, it is $5\times10^{-4}$ and $10^{-2}$ on TwiBot-22 and TwiBot-20, respectively. During the model training process, early stopping techniques and dropout~\cite{srivastava2014dropout} are employed to avoid overfitting.

We apply grid search to adjust the hyperparameters of the LGB model to get the best model configuration for account detection. Specifically, the GNN part of the model adopts two hidden layers, each of which has $512$ channels. During GNN pre-training, the temperature parameter $\tau$ is $0.4$, and on TwiBot-22, the probabilities of dropping edges for the two views are $0.2$ and $0.4$, and on TwiBot-20, they are $0.4$ and $0.6$. Vicuna-7B-v1.5 and ChatGLM3-6B adopt the pre-trained model parameters published in \cite{zheng2023judging} and \cite{du2022glm,zeng2022glm}, respectively, and test them in the zero-shot setting. Model configurations for other baselines follow previous work~\cite{zhou2023detecting,lim2021large,zhou2023semi}. All experiments are performed on NVIDIA A100 80GB GPU, where PyTorch~\cite{paszke2019pytorch} and PyTorch Geometric~\cite{Fey/Lenssen/2019} are used in the experimental implementation.

\vspace{-0.1cm}
\subsection{Overall Results}

For comparative experiments, each experiment is run five times with random weight initialization. The mean and standard deviation $\left (  \mathrm{mean}\pm \mathrm{std}\%\right ) $ on the test set are then calculated and presented in Table~\ref{table:overall_results}. From the experimental results, we have the following observations and discussions:

(1) From the experimental results, our models LGB (GCN) and LGB (GIN) achieve the best and second-best detection results, respectively. Additionally, the test results on two datasets show that our models have achieved significant performance improvement. Specifically, our model LGB (GCN) improves the detection accuracy by $10.95\%$ and $9.98\%$ compared with SIRAN and MixHop, the best baseline models on TwiBot-20 and TwiBot-22, respectively.

(2) For the general and advanced GNNs, our model LGB (GCN) outperforms the best baselines among them on TwiBot-22 and TwiBot-20 by $9.98\%$ and $11.08\%$ in accuracy, respectively. This improvement is attributed to the enhanced semantic information that helps the model effectively detect sparsely linked nodes, thereby improving detection performance.

(3) For the two dedicated social bot detection models, our model LGB (GCN) can still achieve a large improvement in detection accuracy, that is, more than $13.99\%$ and $10.95\%$ on TwiBot-22 and TwiBot-20 respectively, which indicates the effectiveness of the fusion of network structure and node semantics for the social bot detection task.

(4) For the two popular LLMs, our model LGB (GCN) achieves significant accuracy improvements of over $58.49\%$ and $58.63\%$ on TwiBot-22 and TwiBot-20, respectively, which demonstrates the effectiveness of the fusion of structural information with enhanced semantics for detection performance improvement, which we will further verify in Section~\ref{sec:Ablation_Study}.

All these observations above suggest that our model can achieve a great improvement in the social bot detection task by effectively fusing structural and semantic information.

\begin{table*}[t!]
\caption{Overall results. \textbf{Bold} and \underline{underline} represent the best and second best performance, respectively.} 
\centering
\begin{tabular}{@{}ccccccc@{}}
\toprule
{\color[HTML]{000000} \textbf{Datasets}}                    & \multicolumn{3}{c}{{\color[HTML]{000000} \textbf{TwiBot-22}}}                                                                                                   & \multicolumn{3}{c}{{\color[HTML]{000000} \textbf{TwiBot-20}}}                                                                              \\ \midrule
\multicolumn{1}{c|}{{\color[HTML]{000000} \textbf{Method}}} & {\color[HTML]{000000} \textbf{Accuracy}}     & {\color[HTML]{000000} \textbf{F1-Score}}     & \multicolumn{1}{c|}{{\color[HTML]{000000} \textbf{ROC-AUC}}}      & {\color[HTML]{000000} \textbf{Accuracy}}     & {\color[HTML]{000000} \textbf{F1-Score}}     & {\color[HTML]{000000} \textbf{ROC-AUC}}      \\ \midrule
\multicolumn{1}{c|}{{\color[HTML]{000000} GCN}}             & {\color[HTML]{000000} 49.96 ± 0.00}          & {\color[HTML]{000000} 66.63 ± 0.00}          & \multicolumn{1}{c|}{{\color[HTML]{000000} 50.00 ± 0.00}}          & {\color[HTML]{000000} 57.80 ± 0.00}          & {\color[HTML]{000000} 73.26 ± 0.00}          & {\color[HTML]{000000} 50.00 ± 0.00}          \\
\multicolumn{1}{c|}{{\color[HTML]{000000} GIN}}             & {\color[HTML]{000000} 68.42 ± 1.67}          & {\color[HTML]{000000} 68.93 ± 2.80}          & \multicolumn{1}{c|}{{\color[HTML]{000000} 68.40 ± 1.66}}          & {\color[HTML]{000000} 71.79 ± 1.00}          & {\color[HTML]{000000} 77.30 ± 0.56}          & {\color[HTML]{000000} 71.73 ± 0.96}          \\ \midrule
\multicolumn{1}{c|}{{\color[HTML]{000000} GCNII}}           & {\color[HTML]{000000} 68.91 ± 0.11}          & {\color[HTML]{000000} 68.83 ± 0.54}          & \multicolumn{1}{c|}{{\color[HTML]{000000} 68.90 ± 0.11}}          & {\color[HTML]{000000} 76.60 ± 0.51}          & {\color[HTML]{000000} 80.86 ± 0.52}          & {\color[HTML]{000000} 75.75 ± 0.72}          \\
\multicolumn{1}{c|}{{\color[HTML]{000000} GPR-GNN}}          & {\color[HTML]{000000} 72.51 ± 0.15}          & {\color[HTML]{000000} 74.63 ± 0.09}          & \multicolumn{1}{c|}{{\color[HTML]{000000} 72.53 ± 0.14}}          & {\color[HTML]{000000} 76.18 ± 0.67}          & {\color[HTML]{000000} 80.00 ± 0.27}          & {\color[HTML]{000000} 75.41 ± 0.52}          \\
\multicolumn{1}{c|}{{\color[HTML]{000000} MixHop}}          & {\color[HTML]{000000} 73.12 ± 0.09}          & {\color[HTML]{000000} 75.04 ± 0.20}          & \multicolumn{1}{c|}{{\color[HTML]{000000} 73.11 ± 0.09}}          & {\color[HTML]{000000} 76.11 ± 0.91}          & {\color[HTML]{000000} 80.41 ± 0.59}          & {\color[HTML]{000000} 75.66 ± 1.14}          \\
\multicolumn{1}{c|}{{\color[HTML]{000000} APPNP}}           & {\color[HTML]{000000} 62.51 ± 1.41}          & {\color[HTML]{000000} 58.78 ± 8.28}          & \multicolumn{1}{c|}{{\color[HTML]{000000} 62.49 ± 1.42}}          & {\color[HTML]{000000} 65.13 ± 4.17}          & {\color[HTML]{000000} 74.77 ± 0.88}          & {\color[HTML]{000000} 61.55 ± 6.62}          \\
\multicolumn{1}{c|}{{\color[HTML]{000000} LINKX}}           & {\color[HTML]{000000} 72.00 ± 0.06}          & {\color[HTML]{000000} 74.47 ± 0.27}          & \multicolumn{1}{c|}{{\color[HTML]{000000} 72.01 ± 0.06}}          & {\color[HTML]{000000} 62.88 ± 2.12}          & {\color[HTML]{000000} 73.40 ± 0.94}          & {\color[HTML]{000000} 60.59 ± 1.80}          \\
\multicolumn{1}{c|}{{\color[HTML]{000000} H2GCN}}           & {\color[HTML]{000000} 72.24 ± 0.19}          & {\color[HTML]{000000} 74.55 ± 0.20}          & \multicolumn{1}{c|}{{\color[HTML]{000000} 72.27 ± 0.20}}          & {\color[HTML]{000000} 75.96 ± 0.55}          & {\color[HTML]{000000} 81.02 ± 0.56}          & {\color[HTML]{000000} 75.30 ± 0.42}          \\ 
\midrule
\multicolumn{1}{c|}{{\color[HTML]{000000} SIRAN}}           & {\color[HTML]{000000} 70.42 ± 0.06}          & {\color[HTML]{000000} 71.42 ± 0.21}          & \multicolumn{1}{c|}{{\color[HTML]{000000} 70.49 ± 0.15}}          & {\color[HTML]{000000} 76.69 ± 0.75}          & {\color[HTML]{000000} 80.69 ± 0.63}          & {\color[HTML]{000000} 75.83 ± 0.77}          \\
\multicolumn{1}{c|}{{\color[HTML]{000000} CBD (GCN)}}        & {\color[HTML]{000000} 68.58 ± 0.34}          & {\color[HTML]{000000} 69.40 ± 0.52}          & \multicolumn{1}{c|}{{\color[HTML]{000000} 68.58 ± 0.34}}          & {\color[HTML]{000000} 68.78 ± 2.61}          & {\color[HTML]{000000} 70.08 ± 0.61}          & {\color[HTML]{000000} 68.78 ± 2.64}          \\
\multicolumn{1}{c|}{{\color[HTML]{000000} CBD (GIN)}}        & {\color[HTML]{000000} 70.55 ± 0.38}          & {\color[HTML]{000000} 71.33 ± 0.51}          & \multicolumn{1}{c|}{{\color[HTML]{000000} 70.55 ± 0.37}}          & {\color[HTML]{000000} 76.48 ± 1.58}          & {\color[HTML]{000000} 77.20 ± 1.55}          & {\color[HTML]{000000} 77.41 ± 2.20}          \\ 
\midrule
\multicolumn{1}{c|}{{\color[HTML]{000000} Vicuna-7B-v1.5}}           & {\color[HTML]{000000} 50.74 ± 0.18}          & {\color[HTML]{000000} 21.98 ± 0.22}          & \multicolumn{1}{c|}{{\color[HTML]{000000} 50.39 ± 0.18}}          & {\color[HTML]{000000} 47.56 ± 0.78}          & {\color[HTML]{000000} 50.93 ± 0.81}          & {\color[HTML]{000000} 47.31 ± 0.78}          \\
\multicolumn{1}{c|}{{\color[HTML]{000000} ChatGLM3-6B}}           & {\color[HTML]{000000} 49.60 ± 0.20}          & {\color[HTML]{000000} 66.01 ± 0.04}          & \multicolumn{1}{c|}{{\color[HTML]{000000} 49.93 ± 0.23}}          & {\color[HTML]{000000} 53.64 ± 0.17}          & {\color[HTML]{000000} 69.73 ± 0.14}          & {\color[HTML]{000000} 49.62 ± 0.17}          \\
\midrule
\multicolumn{1}{c|}{{\color[HTML]{000000} LGB (GAT)}}          & {\color[HTML]{000000} 80.30 ± 0.06}          & {\color[HTML]{000000} \underline{81.06 ± 0.05}}          & \multicolumn{1}{c|}{{\color[HTML]{000000} 80.30 ± 0.06}}          & {\color[HTML]{000000} 84.83 ± 0.52}          & {\color[HTML]{000000} \underline{87.33 ± 0.47}}          & {\color[HTML]{000000} 83.79 ± 0.53}          \\
\multicolumn{1}{c|}{{\color[HTML]{000000} LGB (GIN)}}    & {\color[HTML]{000000} \underline{80.33 ± 0.03}} & {\color[HTML]{000000} 80.93 ± 0.03} & \multicolumn{1}{c|}{{\color[HTML]{000000} \underline{80.33 ± 0.03}}} & {\color[HTML]{000000} \underline{84.89 ± 0.68}} & {\color[HTML]{000000} 87.16 ± 0.63} & {\color[HTML]{000000} \underline{84.17 ± 0.63}} \\ 
\multicolumn{1}{c|}{{\color[HTML]{000000} \textbf{LGB (GCN)}}}          & {\color[HTML]{000000} \textbf{80.42 ± 0.05}}          & {\color[HTML]{000000} \textbf{81.31 ± 0.08}}          & \multicolumn{1}{c|}{{\color[HTML]{000000} \textbf{80.42 ± 0.05}}}          & {\color[HTML]{000000} \textbf{85.09 ± 0.51}}          & {\color[HTML]{000000} \textbf{87.44 ± 0.31}}          & {\color[HTML]{000000} \textbf{84.23 ± 0.71}}          \\ 
\bottomrule
\end{tabular}
\label{table:overall_results}
\vspace{-0.4cm}
\end{table*}

\vspace{-0.3cm}
\subsection{Online Smart feedback Study} 
\label{section:Online_Smart_feedback_Study}
\begin{figure}[htbp]
\vspace{-0.1cm}
\centering
\setlength{\abovecaptionskip}{0.0cm}   
\includegraphics[width=0.5\linewidth]{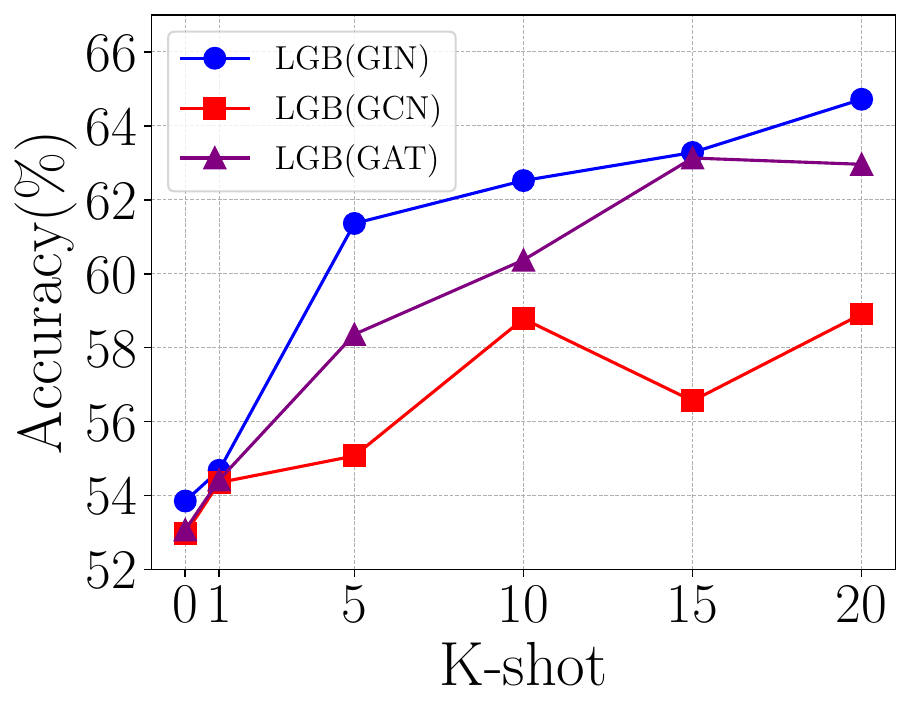}
\caption{Online smart feedback study. X-axis: the number of random samples from each category in the TwiBot-22 training set; Y-axis: the detection accuracy of the model on the TwiBot-22 test set.} \label{fig:online_feedback_acc}
\vspace{-0.6cm}
\end{figure}

To verify the effective detection of new social bots by our model with the assistance of the online smart feedback function, we conduct the following experiments. Specifically, we first prepare the LGB model trained on TwiBot-20, then randomly select only $K$ samples of each category from the TwiBot-22 training set to continue training the LGB model, and then test it on the TwiBot-22 test set. From the experimental results in Fig.~\ref{fig:online_feedback_acc}, we can observe that the detection accuracy of LGB on the TwiBot-22 test set shows a consistent upward trend with the increase of $K$, and there is no sign of slowing down. This should be attributed to the fact that the model learns similar semantic and structural knowledge as in TwiBot-22 during training on TwiBot-20. This knowledge can help the model quickly recognize new social bots. Meanwhile, the online smart feedback function will continuously inject the latest user feedback knowledge into the model, which together ensure the effective detection of constantly evolving bots.

\vspace{-0.3cm}
\subsection{Ablation Study}
\label{sec:Ablation_Study}

To verify the effectiveness of each part of LGB, we conduct ablation studies on TwiBot-22 and TwiBot-20, as follows:

\vspace{-0.1cm}
\begin{itemize}[leftmargin=*,itemsep=0pt,parsep=0.5em,topsep=0.3em,partopsep=0.3em]
    \setlength{\itemsep}{0pt}
    \setlength{\parsep}{0pt}
    \setlength{\parskip}{0pt}
    \item w/o LM Supervised Fine-tuning (SFT): During the training process of the LGB, no SFT of the LM is performed to align it to the social bot detection task.
    
    \item w/o GNN Fine-tuning: During the training process of the LGB, the fine-tuning operation on the GNN is removed.
    
    \item w/o GNN Pre-training: During the training process of the LGB, the pre-training operation of the GNN is removed.
    
    \item w/o Concat: In the process of the LGB training, no concatenation operation is used to fuse semantic information and network structure information.
    
    \item w/ Average: In the process of the LGB training, the averaging operation is used to replace the concatenation to fuse semantic information and network structure information.
    
    \item w/ Max: In the process of the LGB training, the maximum operation is used to replace the concatenation operation to fuse multimodal information.

    \item w/ Supervised Fine-tuned LM: Only the supervised fine-tuned LM (i.e., RoBERTa) in the LGB model is used for account detection.

\end{itemize}

\vspace{-0.1cm}
From the results of the ablation experiments in Table~\ref{table:ablation_study}, the following observations can be obtained:

\begin{table}[t!]
\vspace{-0.4cm}
\caption{Accuracy of ablation experiments.} 
\centering
\begin{tabular}{@{}ccc@{}}
\toprule
{\color[HTML]{000000} \textbf{Ablation Settings}}    & {\color[HTML]{000000} \textbf{TwiBot-22}}    & {\color[HTML]{000000} \textbf{TwiBot-20}}    \\ 
\midrule
{\color[HTML]{000000} \textbf{LGB (GCN) (full model)}}  & {\color[HTML]{000000} \textbf{80.42 ± 0.05}} & {\color[HTML]{000000} \textbf{85.09 ± 0.51}} \\ 
{\color[HTML]{000000} w/o LM Supervised Fine-tuning}  & {\color[HTML]{000000} 69.19 ± 0.43}          & {\color[HTML]{000000} 67.71 ± 2.16}          \\
{\color[HTML]{000000} w/o GNN Fine-tuning}           & {\color[HTML]{000000} 51.15 ± 12.28}          & {\color[HTML]{000000} 54.74 ± 7.01}         \\
{\color[HTML]{000000} w/o GNN Pre-training}          & {\color[HTML]{000000} 80.26 ± 0.22}          & {\color[HTML]{000000} 84.83 ± 0.41}          \\
{\color[HTML]{000000} w/o Concat}                    & {\color[HTML]{000000} 55.75 ± 12.77}          & {\color[HTML]{000000} 72.76 ± 1.98}          \\
{\color[HTML]{000000} w/ Average}                    & {\color[HTML]{000000} 56.04 ± 13.67}          & {\color[HTML]{000000} 85.08 ± 0.49}          \\
{\color[HTML]{000000} w/ Max}                        & {\color[HTML]{000000} 55.95 ± 13.74}          & {\color[HTML]{000000} 84.53 ± 0.57}          \\ 
\midrule
{\color[HTML]{000000} \textbf{LGB (GIN) (full model)}}  & {\color[HTML]{000000} \textbf{80.33 ± 0.03}} & {\color[HTML]{000000} \textbf{84.89 ± 0.68}} \\ 
{\color[HTML]{000000} w/o LM Supervised Fine-tuning}  & {\color[HTML]{000000} 70.45 ± 0.19}          & {\color[HTML]{000000} 75.64 ± 1.73}          \\
{\color[HTML]{000000} w/o GNN Fine-tuning}           & {\color[HTML]{000000} 44.61 ± 6.05}          & {\color[HTML]{000000} 43.36 ± 10.79}         \\
{\color[HTML]{000000} w/o GNN Pre-training}          & {\color[HTML]{000000} 80.02 ± 0.11}          & {\color[HTML]{000000} 84.55 ± 0.13}          \\
{\color[HTML]{000000} w/o Concat}                    & {\color[HTML]{000000} 79.88 ± 0.22}          & {\color[HTML]{000000} 84.00 ± 0.40}          \\
{\color[HTML]{000000} w/ Average}                    & {\color[HTML]{000000} 80.08 ± 0.23}          & {\color[HTML]{000000} 84.10 ± 0.70}          \\
{\color[HTML]{000000} w/ Max}                        & {\color[HTML]{000000} 79.91 ± 0.29}          & {\color[HTML]{000000} 84.47 ± 0.58}          \\ 
\midrule
{\color[HTML]{000000} \textbf{LGB (GAT) (full model)}}  & {\color[HTML]{000000} \textbf{80.30 ± 0.06}} & {\color[HTML]{000000} \textbf{84.83 ± 0.52}} \\ 
{\color[HTML]{000000} w/o LM Supervised Fine-tuning}  & {\color[HTML]{000000} 67.69 ± 0.51}          & {\color[HTML]{000000} 68.89 ± 1.16}          \\
{\color[HTML]{000000} w/o GNN Fine-tuning}           & {\color[HTML]{000000} 48.68 ± 2.90}          & {\color[HTML]{000000} 42.12 ± 0.92}         \\
{\color[HTML]{000000} w/o GNN Pre-training}          & {\color[HTML]{000000} 80.24 ± 0.09}          & {\color[HTML]{000000} 84.72 ± 0.67}          \\
{\color[HTML]{000000} w/o Concat}                    & {\color[HTML]{000000} 55.81 ± 13.24}          & {\color[HTML]{000000} 77.58 ± 1.20}          \\
{\color[HTML]{000000} w/ Average}                    & {\color[HTML]{000000} 55.98 ± 13.58}          & {\color[HTML]{000000} 84.44 ± 0.41}          \\
{\color[HTML]{000000} w/ Max}                        & {\color[HTML]{000000} 56.01 ± 13.58}          & {\color[HTML]{000000} 84.64 ± 0.28}          \\ 
\midrule
{\color[HTML]{000000} w/ Supervised Fine-tuned LM}  & {\color[HTML]{000000} 80.01 ± 0.15}          & {\color[HTML]{000000} 84.70 ± 0.22}          \\
\bottomrule

\end{tabular}
\label{table:ablation_study}
\vspace{-0.6cm}
\end{table}

(1) Replacing or removing any component of the full model results in performance degradation, demonstrating that each component contributes to the effectiveness of LGB.

(2) From w/o LM Supervised Fine-tuning, we can see that using the LM without supervised fine-tuning during the training of the GNN part significantly degrades the performance of the model, i.e., the detection accuracy on TwiBot-22 and TwiBot-20 decreases by $12.30\%-15.70\%$ and $10.90\%-20.43\%$, respectively. This decline is attributed to the numerous isolated and less-linked nodes in the network. Without the injection of enhanced semantic information, the LGB model cannot distinguish them effectively. This further verifies the importance of fusing node semantics and network structure for account detection tasks.

(3) From w/o GNN Fine-tuning, we observe that without fine-tuning the GNN, the model performance decreases greatly, i.e., the performance drops by $36.40\%-44.47\%$ and $35.67\%-50.35\%$ on TwiBot-22 and TwiBot-20, respectively. This decline occurs because the semantic information input by the LM and the structural information extracted by the GNN are not fused and aligned to the account detection task through fine-tuning. Therefore, the effective fusion of multi-modal information is crucial for improving the model performance. Meanwhile, we further explore it in the following experiments w/o Concat, w/ Average, and w/ Max.

(4) From w/o GNN Pre-training, removing the pre-training stage from the GNN training process will cause the model performance to degrade, which proves that it has a certain contribution to the model detection performance. Moreover, our label robustness experiments in Section~\ref{sec:Robustness_Study} further verify the valuable knowledge learned from unlabeled data in the GNN pre-training stage helps reduce the model's dependence on labels to achieve performance improvement in the case of extreme lack of labels.

(5) From w/o Concat, w/ Average, and w/ Max, we observe that removing or replacing the concat operation in the GNN fine-tuning stage results in varying degrees of degradation in the performance of the model, illustrating the effectiveness of the concat operation for multi-modal information fusion.

(6) From w/ Supervised Fine-tuned LM, we find that removing the GNN part leads to a certain degradation of the model performance, indicating that the network structure plays a role in improving the detection accuracy. In addition, we will further verify in Section~\ref{sec:Robustness_Study} that when the node text information is destroyed, the model can enhance its detection performance by fully mining the relationship information contained in the network structure.

\begin{figure*}[t!]
\centering
\subfloat[]{\includegraphics[width=0.32\linewidth]{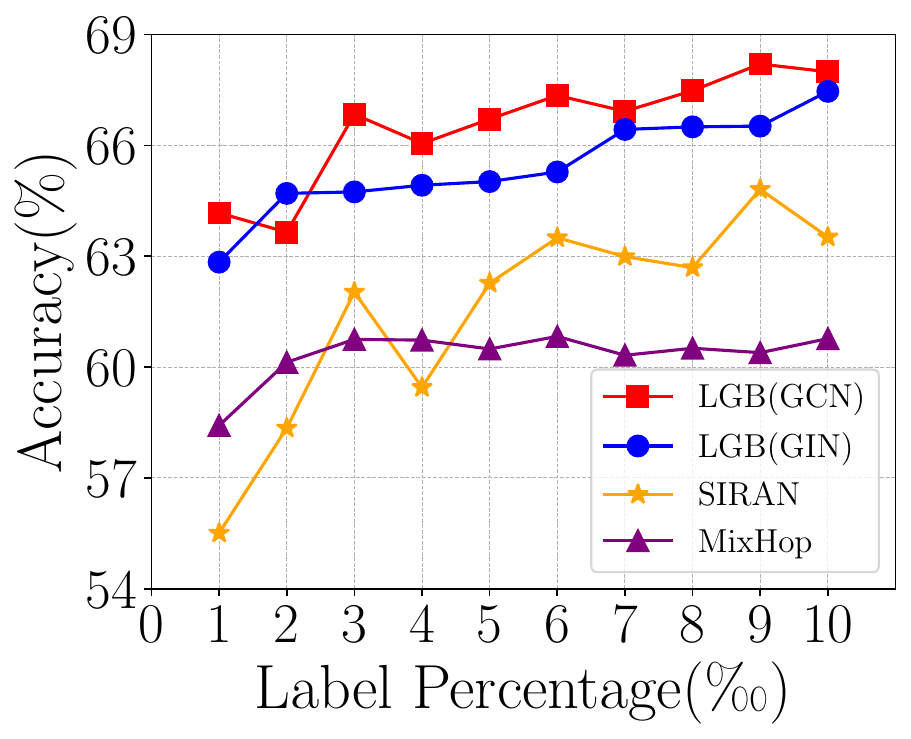}%
\label{subfig:label_robustness_study}}
\hfil
\centering
\subfloat[]{\includegraphics[width=0.32\linewidth]{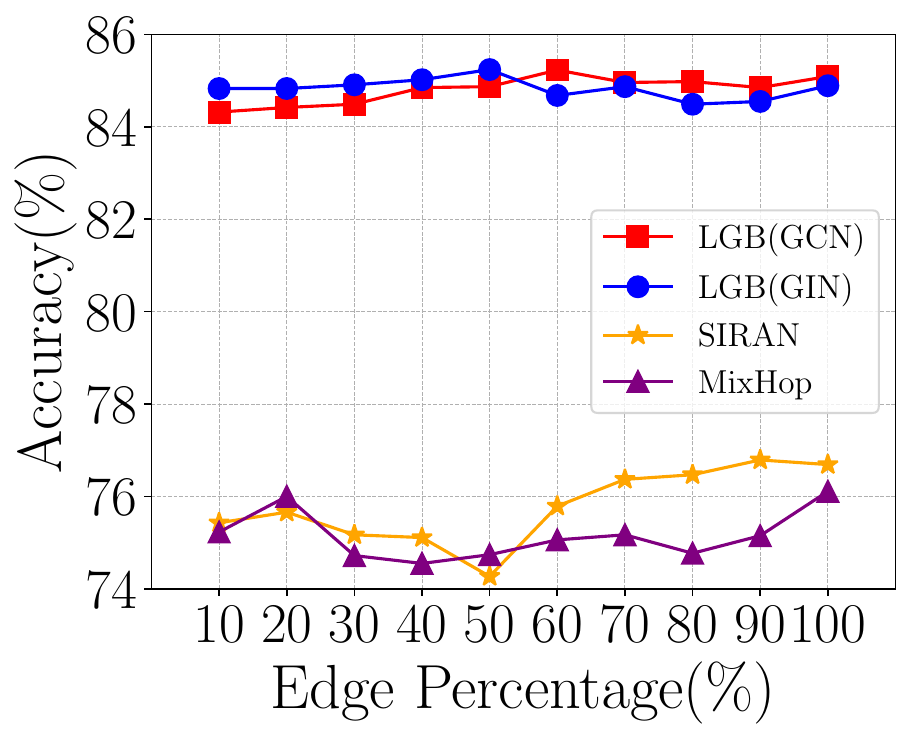}%
\label{subfig:edge_robustness_study}}
\hfil
\centering
\subfloat[]{\includegraphics[width=0.32\linewidth]{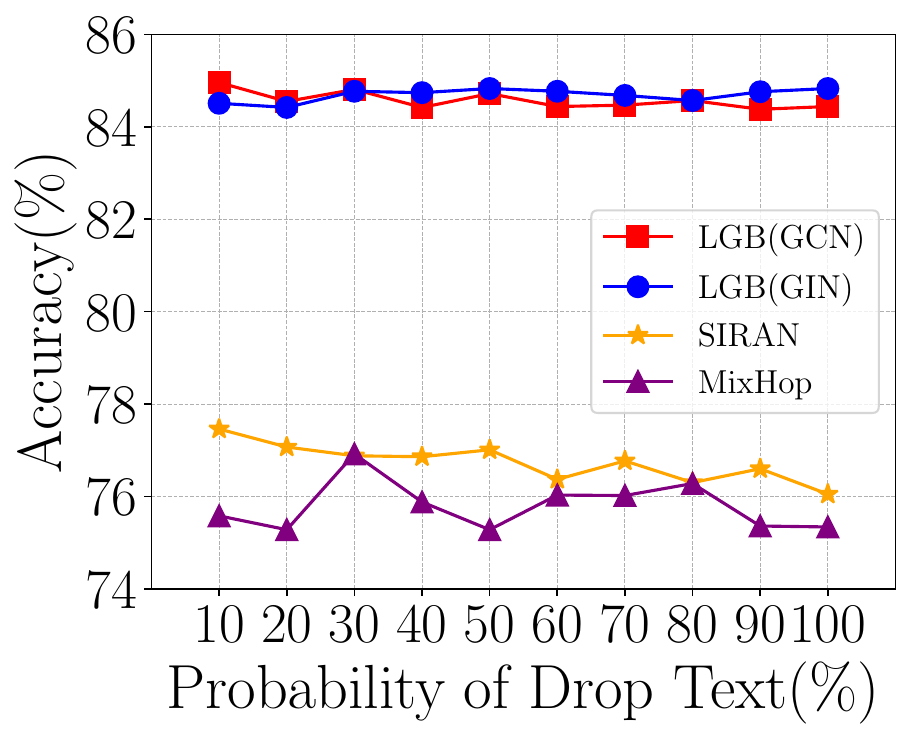}%
\label{subfig:feature_robustness_study}}
\caption{Robustness study. Y-axis: the detection accuracy of the model on the TwiBot-20 test set; X-axis: (a) the proportion of sample labels, (b) the proportion of edges sampled from the TwiBot-20 training set, and (c) the probability of randomly dropping $10\%$ of each user's text sequence.}
\label{fig:robustness_study}
\vspace{-0.6cm}
\end{figure*}

\vspace{-0.2cm}
\subsection{Robustness Study}
\label{sec:Robustness_Study}

Existing social bot detection models usually rely on a large number of high-quality labeled data. However, due to the high cost of data acquisition and manual annotation and the bots' rapid evolution, these needs cannot be met, resulting in degraded detector performance. Given this, we select the best two LGB models to carry out robustness studies on TwiBot-20 to evaluate their robustness.

\noindent \textbf{Label robustness study.} Firstly, to simulate the scenario where labels are scarce to verify the robustness of the model, we only randomly sample $0.1\%-1\%$ of the labels from the training set for model training and then test it on the test set. The experimental results are presented in Fig.~\ref{subfig:label_robustness_study}. 
\textbf{Edge robustness study.} Secondly, given that bots evade detection by establishing social relationships with humans, to verify the robustness of the model for the network structure, we randomly sample $10\%-100\%$ of the edges (i.e., social relationships) from the network to carry out the edge robustness studies. The experimental results are shown in Fig.~\ref{subfig:edge_robustness_study}. \textbf{Feature robustness study.} Thirdly, considering that the creators of social bots deliberately miss or forge account attribute information to increase the difficulty of detection, we conduct feature robustness studies to verify the robustness of the model to perturbations of input user text information. Specifically, for each user's text sequence in the training set (shown in Fig.~\ref{fig:user_text_sequence}), we randomly remove $10\%$ of the sequence with a probability of $10\%-100\%$. Then we train the model using these corrupted sequences and evaluate it on the test set. The results are shown in Fig.~\ref{subfig:feature_robustness_study}.

\noindent \textbf{Analysis and discussion.} Based on the experimental results in Fig.~\ref{fig:robustness_study}, we have the following analyses and discussions: (1) From the results of robustness experiments, it can be seen that our model consistently outperforms the baseline models, which proves that our model is more robust. (2) For the label robustness study, when the training sample labels are extremely scarce, that is, $0.1\%-1\%$ of the training labels, the performance of LGB can still consistently exceed that of the baselines by a large margin (more than $2.64\%$) and shows a trend of continuous growth. This is due to the valuable knowledge learned by the GNN model from unlabeled data in the pre-training stage, which helps the model reduce its dependence on labels and achieve performance improvements with fewer labels. (3) For the edge robustness study, our model always outperforms the baseline models by a large margin (more than $10.11\%$) with various edge percentages, and the performance is more stable. This is because when the number of edges is small, the LM can effectively extract semantic knowledge from node text sequences to enhance detection performance. (4) For the feature robustness study, the performance of LGB is consistently better than that of the baselines by a large margin (more than $9.10\%$), and the performance is more stable. This is because when the textual sequences of nodes are destroyed, the GNN model can still ensure effective detection of social accounts by extracting structural information from the network.

\vspace{-0.1cm}
\section{Related Work of Social Bot Detection}
\label{sec:related_work}

Based on the way social data is used, social bot detection can be divided into the following three categories: feature-based, content-based, and graph-based approaches.

\noindent \textbf{Feature-based approaches:} Crowdsourcing and statistical learning approaches based on feature engineering are employed for account detection by extracting features such as attribute information~\cite{yang2020scalable}, tweets~\cite{heidari2021empirical,wang2018social}, and social behaviors~\cite{sayyadiharikandeh2020detection,cai2017behavior} of users. Recent work focuses on improving the performance of feature-based detectors through feature selection~\cite{mbona2022feature}, multiple feature fusion~\cite{fazil2021deepsbd}, and extraction of balanced distribution features~\cite{li2022novel}. 
However, such methods are vulnerable to feature forgery attacks~\cite{cresci2020decade,latah2020detection}.

\noindent \textbf{Content-based approaches:} Compared with feature-based approaches, inspired by the fact that social bots often achieve their malicious purposes through the dissemination of fake tweets, content-based approaches mainly focus on tweet content. Such methods leverage content analysis technology to evaluate the authenticity and intent of tweets for bot detection. For example, bidirectional Long Short-Term Memory (BiLSTM) is used to extract content features from tweets for bot detection~\cite{wei2019twitter}. Heidari et al.~\cite{heidari2020deep} use Embeddings from Language Models (ELMo)~\cite{peters2018deep} to encode users' tweets to obtain better representations. Recent research in this area combines tweet content analysis with information such as user attributes~\cite{fazil2021deepsbd} or geographical location~\cite{ravazzi2022towards} for account detection. However, the recent rapid development of LLM applications, such as ChatGPT, has enhanced the creative capabilities of social bots, posing significant challenges to content-based detection approaches.

\noindent \textbf{Graph-based approaches:} Different from content-based and feature-based approaches, inspired by the important role played by the strength~\cite{bakshy2012role} and structural diversity~\cite{ugander2012structural, zhang2013social} of social relationships in the spread of false information, graph-based approaches treat social accounts as nodes and social relationships as edges to model social networks for bot detection. Specifically, through the message-passing mechanism, this method aggregates the information from neighbors to the central node to enhance the representation capability for stronger detection performance. For example, Ali Alhosseini et al. \cite{ali2019detect} use graph convolutional networks (GCN) to learn low-dimensional representations of nodes for bot detection. Zhou et al. \cite{zhou2023semi} design a semi-supervised initial residual relation attention networks (SIRAN), which improves the model performance by employing a heterophily-aware relation attention strategy. However, our research reveals that about $55.34\%$ of nodes in social networks are either isolated or have only one neighbor, as shown in Fig.~\ref{fig:Figure1_data_show}. For the detection of these nodes, due to the lack of social relationships, it is impossible to effectively aggregate social information to obtain enhanced node representation, resulting in the performance of traditional graph-based approaches being significantly weakened. These sparsely linked nodes contain hidden bots that will be quickly activated to establish links with humans when performing malicious tasks to spread false information and engage in malicious activities. These covert and harmful bots pose new challenges to the account detection task. Therefore, the main purpose of this work is to explore a more effective detection method for sparsely linked bots in social networks.

\vspace{-0.1cm}
\section{Conclusion}
\label{sec:Conclusion}

In this paper, we focus on the task of social bot detection. By analyzing real-world social network data, we find that there are a large number of isolated and poorly linked nodes, posing a significant challenge to graph-based detection methods. To solve this issue, we propose a novel social bot detection framework LGB, which comprises two main parts: GNN and LM. Specifically, first, the unified user text, constructed from social account information, is fed into the LM for supervised fine-tuning to better understand social account semantics. Then, the node representations encoded by the supervised fine-tuned LM are input into the pre-trained GNN to further enhance them by injecting network structure information. Finally, the LGB model improves its ability for account detection by fusing information from two modalities: node semantics and network structure. Meanwhile, to combat the rapid evolution of bots, at the system architecture level, we design a smart feedback function, enabling the model to evolve continually by incorporating feedback information from online expert users, thereby further enhancing its account detection capabilities. Extensive experiments on two real-world social bot detection benchmarks demonstrate that LGB consistently outperforms state-of-the-art baselines. To better help people identify malicious social bots and promote social safety, we have released LGB online, which receives widespread attention.

\vspace{-0.2cm}
\vpara{Limitation and future work:} Because of the high data acquisition costs and the different distribution of user data across various social platforms, joint detection across multiple platforms remains an open issue in this field, and LGB does not yet support this capability. We will study it in future work.
\vspace{-0.6cm}

\bibliographystyle{IEEEtran}
\bibliography{IEEEabrv,ref}

\vspace{-0.9cm}
\def \bioGapReduce{-5pt}

\begin{IEEEbiography}
[{\includegraphics[width=1in,height=1.25in,clip,keepaspectratio]{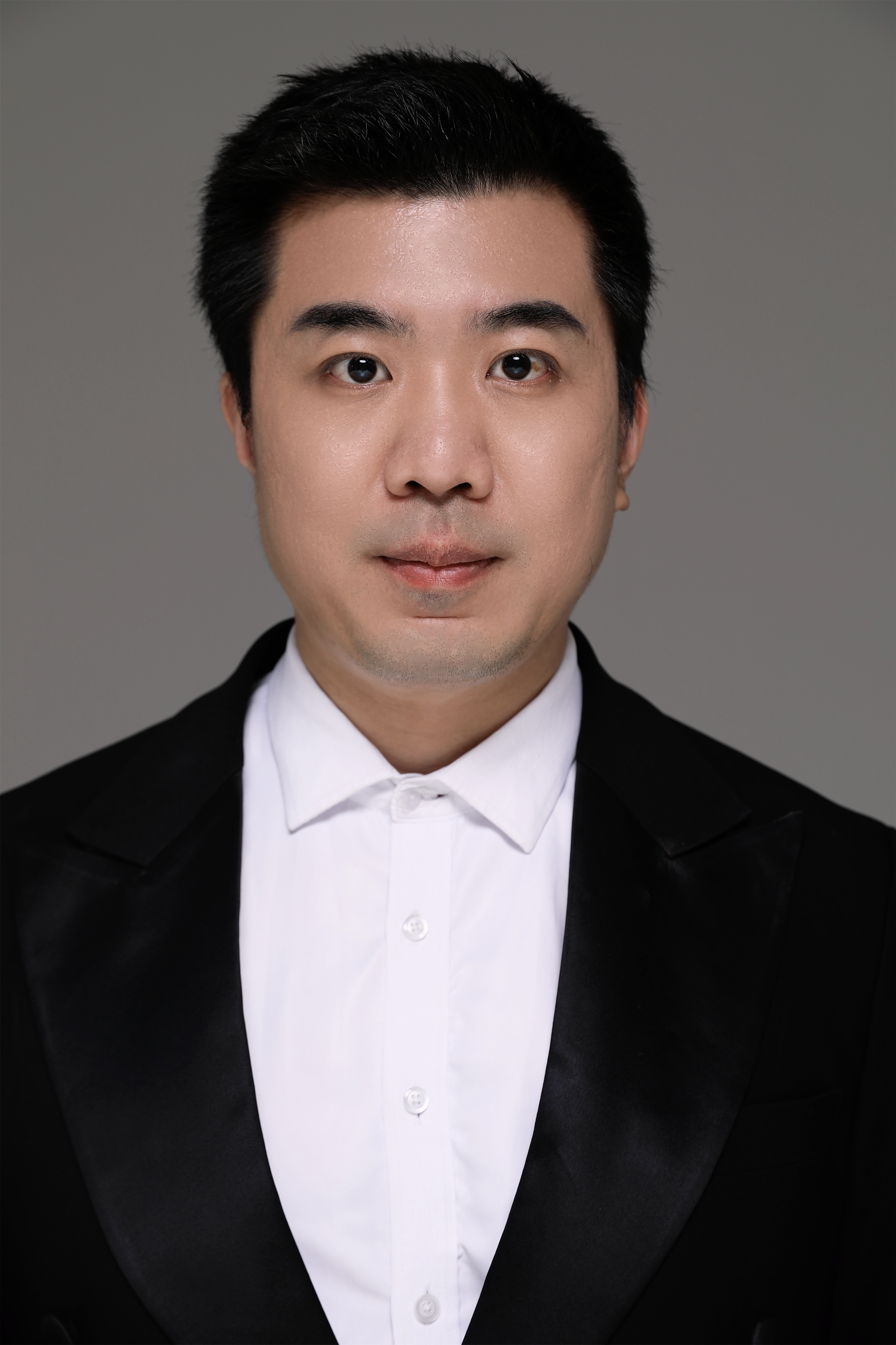}}]
{Ming Zhou} is a PhD candidate in the Department of Computer Science and Technology, Tsinghua University.
Before joining Tsinghua, he worked in research and development at Baidu and Tencent.
His research interests include social networks, language models, and graph representation learning.
He received the 2023 ECML PKDD Best Student Paper Award.
\end{IEEEbiography}
\vspace{\bioGapReduce}
\vspace{-0.8cm}

\begin{IEEEbiography}[{\includegraphics[width=1in,height=1.25in,clip,keepaspectratio]{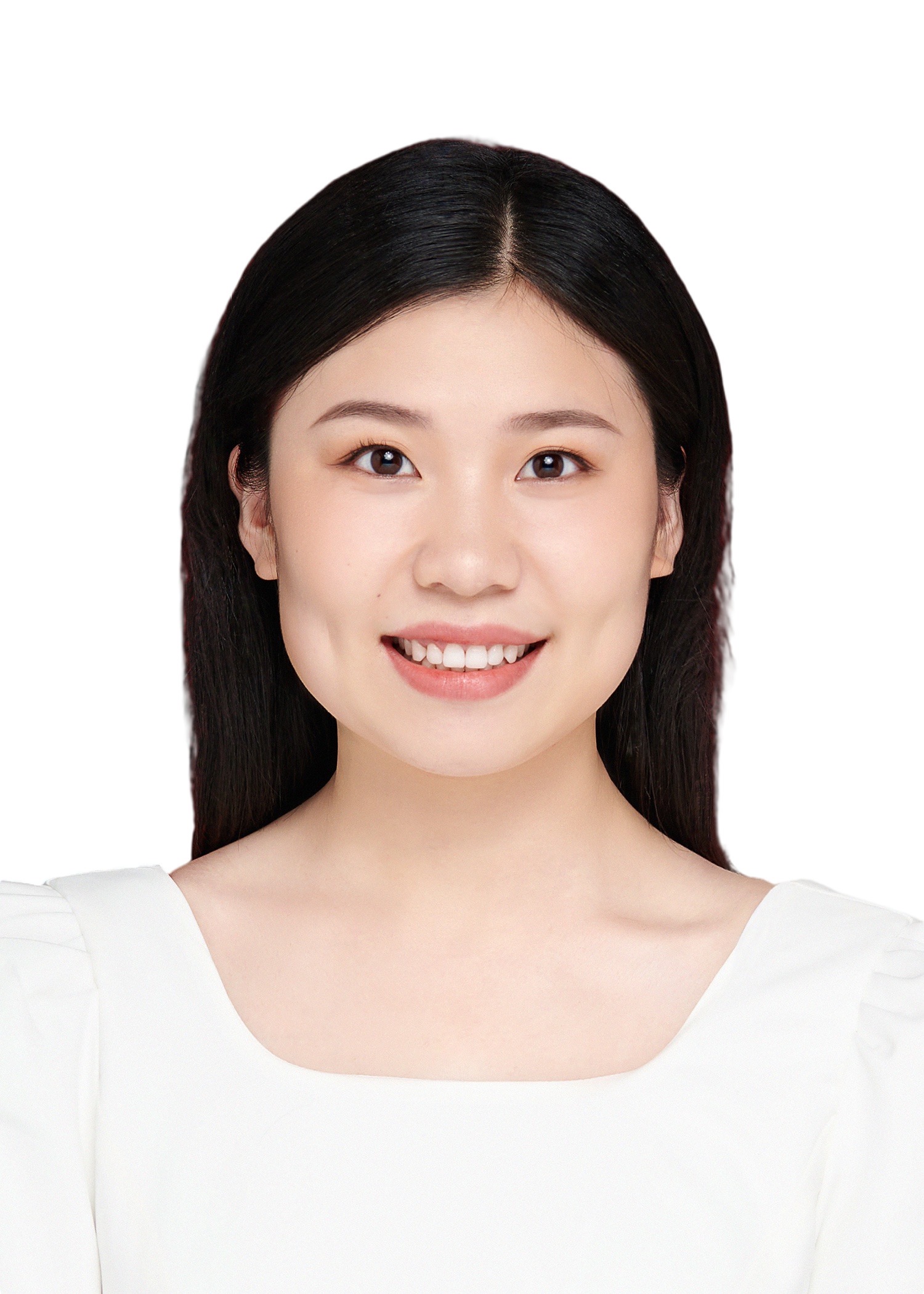}}]{Dan Zhang} is a PhD candidate in the Department of Computer Science and Technology, Tsinghua University. 
She got her master's degree from Software of School, Tsinghua University. 
Her research interests include language models, graph representation learning, and recommendation systems.
She has published papers at top conferences and journals, such as WWW, TKDE, KDD, etc.
\end{IEEEbiography}
\vspace{\bioGapReduce}
\vspace{-0.6cm}

\begin{IEEEbiography}[{\includegraphics[width=1in,height=1.25in,clip,keepaspectratio]{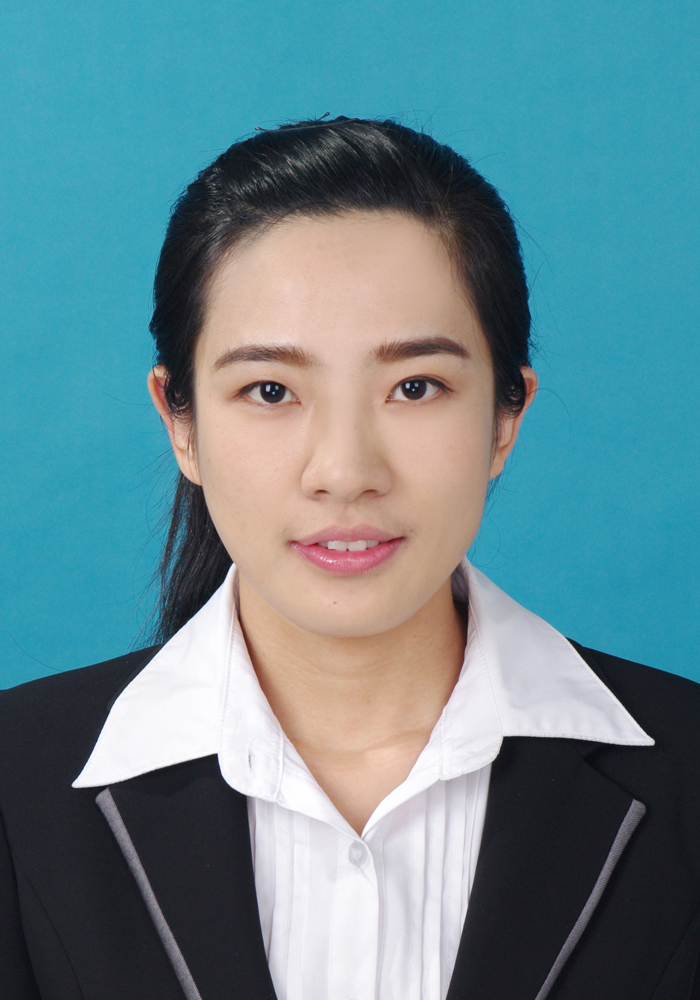}}]{Yuandong Wang} is currently an assistant researcher with the  Department of Computer Science and Technology, Tsinghua University. She received her PhD degree in computer science from Beihang University. Her research interests include natural language understanding, spatiotemporal prediction, applications of pre-trained models, and graph neural networks in cross-domains. She has more than 10 papers published in top international conferences and journals, such as KDD, ICDE, TKDE, etc.
\end{IEEEbiography}
\vspace{\bioGapReduce}
\vspace{-0.8cm}

\begin{IEEEbiography}[{\includegraphics[width=1in,height=1.25in,clip,keepaspectratio]{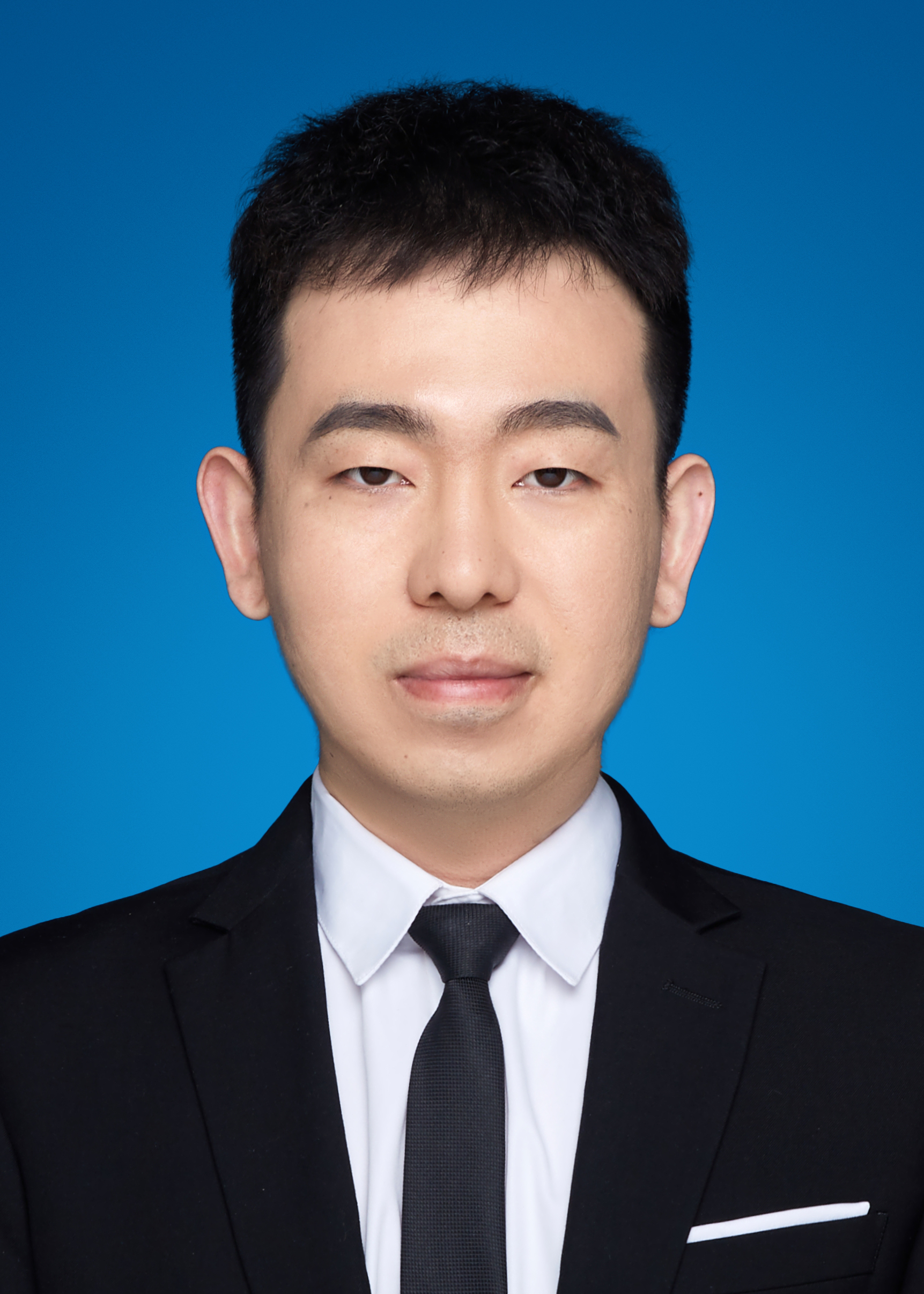}}]{Yangli-ao Geng} received his Ph.D. in Computer Science and Technology from Beijing Jiaotong University, Beijing, China, in 2021. He is currently an assistant Professor at Beijing Jiaotong University. His research interests include graph neural networks and spatiotemporal data mining.
\end{IEEEbiography}
\vspace{\bioGapReduce}
\vspace{-0.8cm}

\begin{IEEEbiography}[{\includegraphics[width=1in,height=1.25in,clip,keepaspectratio]{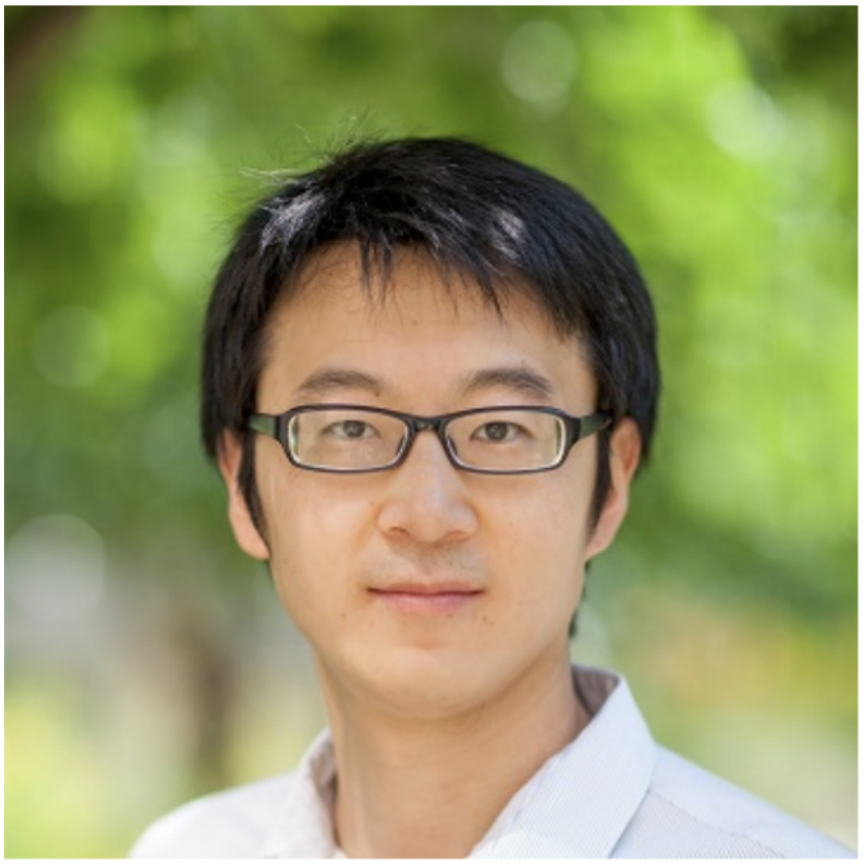}}]{Yuxiao Dong} is an assistant professor of computer science at Tsinghua University. His research focuses on data mining, graph representation learning, pre-training models, and social networks, with an emphasis on developing machine learning models to addressing problems in Web-scale systems. He received the 2017 SIGKDD Dissertation Award Honorable Mention and 2022 SIGKDD Rising Star Award.
\end{IEEEbiography}
\vspace{-0.8cm}

\begin{IEEEbiography}[{\includegraphics[width=1in,height=1.25in,clip,keepaspectratio]{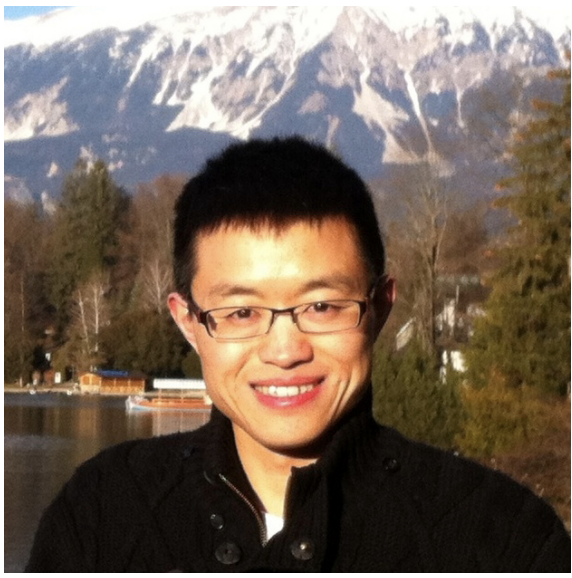}}]{Jie Tang} is a Professor of the Department of Computer Science and Technology at Tsinghua University. His interests include data mining, social networks, and machine learning. He has published over 200 research papers in top international journals and conferences. He served as Associate General Chair of KDD 2018, and acting Editor-in-Chief of ACM TKDD. He was honored with NSFC Distinguished Young Scholar, and 2018 SIGKDD Service Award.
\end{IEEEbiography}

\vfill

\end{document}